\documentclass[%
reprint,
 aps,
 prx,
]{revtex4-2}

\usepackage{graphicx}
\graphicspath{{./figs/}}
\usepackage{dcolumn}
\usepackage{hyperref}
\hypersetup
{
pdfmenubar=true, 
pdfhighlight=/O, 
pdfpagelayout=SinglePage, 
pdffitwindow=true, 
colorlinks=true, 
linkcolor=blue, 
citecolor=blue, 
urlcolor=blue 
}

\usepackage{amsmath, amsthm, amssymb, amsfonts, amsbsy}
\usepackage{mathtools} 
\usepackage{bm}

\newcommand{\upi}{\pi}

\DeclareMathOperator*{\argmax}{arg\hspace{0.8pt}max}
\DeclareMathOperator*{\argmin}{arg\hspace{0.8pt}min}
\DeclareMathOperator*{\ceil}{ceil}
\DeclareMathOperator*{\Bayes}{Bayes}
\DeclareMathOperator*{\Mean}{Mean}

\DeclareMathOperator*{\Std}{Std}

\renewcommand*{\v}[1]{\boldsymbol{#1}}

\renewcommand{\d}{\mathrm{d}} 
\newcommand{\id}{\; \d} 

\newcommand{\normtwo}[1]{\lVert #1 \rVert_2}
\newcommand{\normone}[1]{\lVert #1 \rVert_1}

\usepackage[capitalise]{cleveref}
\crefformat{equation}{Eq. (#2#1#3)} 
\Crefformat{equation}{Equation (#2#1#3)} 
\crefrangeformat{equation}{Eqs. (#3#1#4--#5\crefstripprefix{#1}{#2}#6)} 
\Crefrangeformat{equation}{Equations (#3#1#4--#5\crefstripprefix{#1}{#2}#6)} 
\crefmultiformat{equation}{Eqs. (#2#1#3)}{ and~(#2#1#3)}{, (#2#1#3)}{ and~(#2#1#3)} 
\Crefmultiformat{equation}{Equations (#2#1#3)}{ and~(#2#1#3)}{, (#2#1#3)}{ and~(#2#1#3)} 

\usepackage{xcolor}
\newcommand{\revision}[1]{\textcolor{black}{#1}}

\begin{document}

\title{Searching for a source without gradients: how good is infotaxis and how to beat it}

\author{Aurore Loisy}
\email[]{aurore.loisy@irphe.univ-mrs.fr}
\author{Christophe Eloy}
\email[]{christophe.eloy@centrale-marseille.fr}
\affiliation{Aix Marseille Univ, CNRS, Centrale Marseille, IRPHE, Marseille, France}

\date{\today}

\begin{abstract}
Infotaxis is a popular search algorithm designed to track a source of odor in a turbulent environment using information provided by odor detections. To exemplify its capabilities, the source-tracking task was framed as a partially observable Markov decision process consisting in finding, as fast as possible, a stationary target hidden in a 2D grid using stochastic partial observations of the target location. Here we provide an extended review of infotaxis, together with a toolkit for devising better strategies. We first characterize the performance of infotaxis in domains from 1D to 4D. Our results show that, while being suboptimal, infotaxis is reliable (the probability of not reaching the source approaches zero), efficient (the mean search time scales as expected for the optimal strategy), and safe (the tail of the distribution of search times decays faster than any power law, though subexponentially). We then present three possible ways of beating infotaxis, all inspired by methods used in artificial intelligence: tree search, heuristic approximation of the value function, and deep reinforcement learning. The latter is able to find, without any prior human knowledge, the (near) optimal strategy. Altogether, our results provide evidence that the margin of improvement of infotaxis toward the optimal strategy gets smaller as the dimensionality increases.
\end{abstract}


\maketitle

\section{Introduction}

Tracking down a source of odor in a turbulent environment is a task performed by numerous animals \cite{Vickers2000,Hansson1999book,Murlis1992,Koehl2001,Reidenbach2011,Kiorboe2018book,Reddy2022}. In the hope of uncovering the search algorithms used in Nature, Vergassola et al.~\cite{Vergassola2007} formulated what we will refer to as the ``source-tracking problem''. The modeling of this biology-inspired problem relies on physics. Its solutions, however, call for methods of operations research, automated planning and artificial intelligence. Our first objective here is to make this problem and its solutions accessible to these different communities, which are not always aware of each other's work. 

A source that emits a chemical substance, such as an odor, in a quiescent fluid generates a smooth concentration field that decays with the distance to the emission point. This source is easily tracked, because one only needs to follow concentration gradients: this behavior, ubiquitous in biology, is called chemotaxis. Source-tracking becomes much harder in a turbulent medium, because the concentration field consists of disconnected patches of high concentration, randomly distributed and separated by voids. An animal able to detect concentration levels would receive a highly intermittent signal consisting of sharp peaks separated by long periods with no measurable concentration \cite{Celani2014,Reddy2022}. Gradient-based strategies are therefore doomed to fail in turbulence.

To mimic these searching conditions in a computation-friendly environment, Vergassola et al.~\cite{Vergassola2007} designed the source-tracking problem. It consists in minimizing the number of steps (cumulated cost) to reach a source (target) hidden in a discrete grid. At each step, the searcher (agent) moves to a neighbor cell (action) and receives a stochastic sensory signal (observation, called ``hits''). These observations model odor detection events, and provide a noisy information about the distance to the source (stochastic partial information). The agent knows how observations are generated (the model), and has a perfect memory of past observations and actions, therefore it can maintain a probability distribution over source locations (belief) updated after each observation using Bayes' inference theorem. Solving this problem consists in finding the optimal strategy (policy), defined as the mapping from belief to action, that minimizes the expected number of steps to reach the source.

In the language of artificial intelligence and related disciplines, the source-tracking problem is a partially observable Markov decision process (POMDP) which is framed as a belief-MDP (a MDP -- Markov decision process -- where states are replaced by belief states). It actually belongs to a narrower class of problems, called partially observable stochastic shortest path problems, for which a few formal mathematical results exist, namely: the existence of a deterministic policy that is optimal within the class of Markov policies, and the pointwise convergence of value iteration to the unique bounded fixed point of the dynamic programming operator \cite{Patek2001,Patek2007}. 

Yet computing the optimal policy is not feasible: POMDP exact solvers take prohibitively large amounts of computation time for any but the smallest problems \cite{Chatterjee2016,Shani2013,Kochenderfer2022book,Kurniawati2022}, and the continuous nature of the belief space prevents the use of exact methods for tabular MDPs (e.g., value iteration). Approximate solution methods exist \cite{Hauskrecht2000,Ross2008,Kochenderfer2022book,Kurniawati2022}, but scalability remains an issue. An alternative is deep reinforcement learning: originally designed to solve MDPs using a neural network approximation of the value function \cite{Sutton2018book}, existing algorithms can account for partial observability using the belief-MDP formulation.

Another approach is to handcraft heuristic policies using intuition and knowledge. Infotaxis is a such a heuristic: it states that the agent should choose the action from which it expects the greatest information gain about the source location. The physical intuition behind this algorithm is, quoting the authors, that ``information accumulates faster close to the source because cues arrive at a higher rate, hence tracking the maximum rate of information acquisition will guide the searcher to the source much like concentration gradients in chemotaxis'' \cite{Vergassola2007}. 

The infotaxis policy is more precisely defined as the policy that maximizes the expected gain of information, defined as the decrement of Shannon entropy of the belief, over a 1-step horizon. In other words, infotaxis is an information-greedy policy. The idea of information maximization (or, equivalently, uncertainty minimization) using a probabilistic formulation can be traced back to Cassandra et al.~\cite{Cassandra1996} who proposed it as a heuristic for robot navigation in uncertain environments, and it has since become a central concept for robotic exploration algorithms \cite{Thrun2006book}. 

Infotaxis is generally believed to be robust and has become a popular search algorithm. It has been implemented in robots \cite{Moraud2010,Lochmatter2010thesis,Masson2013} and has inspired various extensions: infotaxis on different types of lattices \cite{Masson2009,Rodriguez2014}, continuous-space infotaxis \cite{Barbieri2011}, mapless infotaxis \cite{Masson2013}, collective infotaxis \cite{Masson2009}, socialtaxis \cite{Karpas2017}, entrotaxis \cite{Hutchinson2018}, and energy-constrained proportional betting \cite{Chen2020}. Its relevance to biological searches remains speculative, but some attempts have been made to explain the trajectories of moths \cite{Vergassola2007,Voges2014} and of the worm \textit{C. elegans} \cite{Calhoun2014} by infotactic searches. 

Yet, the supposedly good performance of infotaxis is somewhat surprising: there is, a priori, no reason why greedily reducing uncertainty should minimize the time to reach the source (unless the two were linearly related; it is easy to check that they are not). Besides, infotaxis \revision{was designed to bring the agent close to the source in a robust manner. It is known not to be optimal with respect to the search time,} and that better performance can be achieved by shifting it toward a more exploitative behavior. This latter approach is not satisfactory though, because it introduces a tunable parameter which appropriate value is far from universal \cite{Masson2009,Masson2013}.

\revision{Several papers implemented (variants of) infotaxis and confirmed its ability to lead to the source in various searching conditions \cite{Masson2009,Barbieri2011,Rodriguez2014,Ristic2016,Eggels2017,Rodriguez2017,Hutchinson2018}. A number of studies also evaluated its robustness to model uncertainty \cite{Masson2009,Ristic2016,Rodriguez2017,Hutchinson2018}. Nevertheless, the performance of the original infotaxis algorithm has never been assessed in a \emph{systematic} manner.} In addition, most of these prior studies applied infotaxis to searches in 2D domains: very few considered the 3D problem \cite{Barbieri2011,Eggels2017}, and none tested the algorithm in 1D. The questions of how infotactic trajectories are affected by dimensionality and whether infotaxis generalizes well to high dimensional spaces \revision{remain vastly unexplored}.

In this paper, our main contributions are the following. (1) We generalize the source-tracking problem to any space dimension and we reformulate it using the POMDP framework, which allows us to formally write the optimal strategy as the solution of a recurrence equation (known as the Bellman optimality equation). 
(2) We introduce a protocol for a rigorous evaluation of intofaxis performance (in the sense that it is not affected by finite-size effects or by an arbitrarily chosen initial distance to the source) and we propose an efficient method to compute the distribution of arrival times.
(3) Using this protocol, we compute and analyze the performance of infotaxis as a function of the problem dimensionality (from 1D to 4D) and of the relevant parameters which govern the physics of odor propagation and detection.
(4) We evaluate to which extent three well-established techniques from artificial intelligence are able to improve on infotaxis, or even to yield optimal strategies: (i) tree search based on an existing heuristic (``N-step infotaxis''), (ii) knowledge-based approximation of the optimal value function (``space-aware infotaxis''), and (iii) deep reinforcement learning.

The lack of directionality in the information provided by observations makes the source-tracking problem particularly challenging. \revision{The hardest scenario is the one where no mean flow is present to break the radial symmetry of the problem. We focus on this worst-case scenario, which is expected to provide lower bounds on performance for situations where a mean flow is present, as is often the case in real applications.}
Our results provide extensive evidence that, while being suboptimal, infotaxis has three important properties: (i) reliability (the probability of never finding the source approaches zero), (ii) efficiency (the mean search time scales as expected for the optimal policy), and (iii) safety (in the sense that arrival times are not plagued by large fluctuations). The performance of space-aware infotaxis and of (near) optimal policies obtained by reinforcement learning strongly suggest that while infotaxis is vastly suboptimal in 1D, the margin of improvement toward the optimal policy gets tighter as the dimensionality increases. 

The remainder of this paper is organized as follows. The source-tracking problem and its POMDP formulation are presented in \cref{sec:source-tracking}, together with our policy evaluation protocol. Infotaxis and its generalization to multi-step anticipation are described and evaluated in \cref{sec:infotaxis}. Space-aware infotaxis is introduced in \cref{sec:space_aware_infotaxis}, where it is shown to do better than infotaxis without any tunable parameter. \Cref{sec:RL} is concerned with deep reinforcement learning and presents results for (near) optimal policies. The reasons behind the good performance of infotaxis are discussed in \cref{sec:discussion}. Conclusions are drawn in \cref{sec:conclusion}.

\section{The source-tracking POMDP}
\label{sec:source-tracking}

\subsection{Overview}

\begin{figure*}
    \centering
    \includegraphics[width=0.85\linewidth]{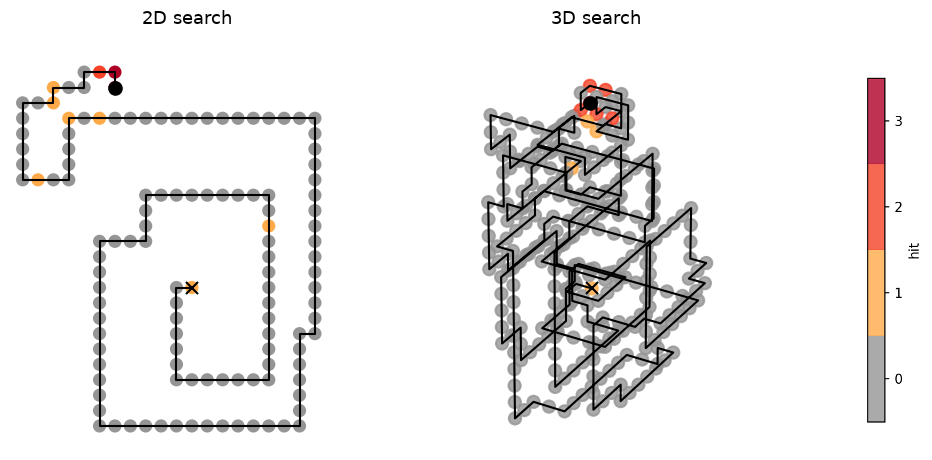}
    \caption{
    Examples of trajectories in the source-tracking problem, where a searcher (agent) must find a source (hidden target) using information provided by odor detections called ``hits'' (partial observations). The start of the search is indicated by a cross, and the end of the search (source location) is depicted by a black dot. These trajectories were generated using the infotaxis policy. Videos are provided in Supplementary Material. 
    \label{fig:illustration_source_tracking_problem}
	}
\end{figure*}

\begin{figure*}
    \flushleft
    \hspace{1cm} (a) \\
    \centering
    \includegraphics[width=0.99\linewidth]{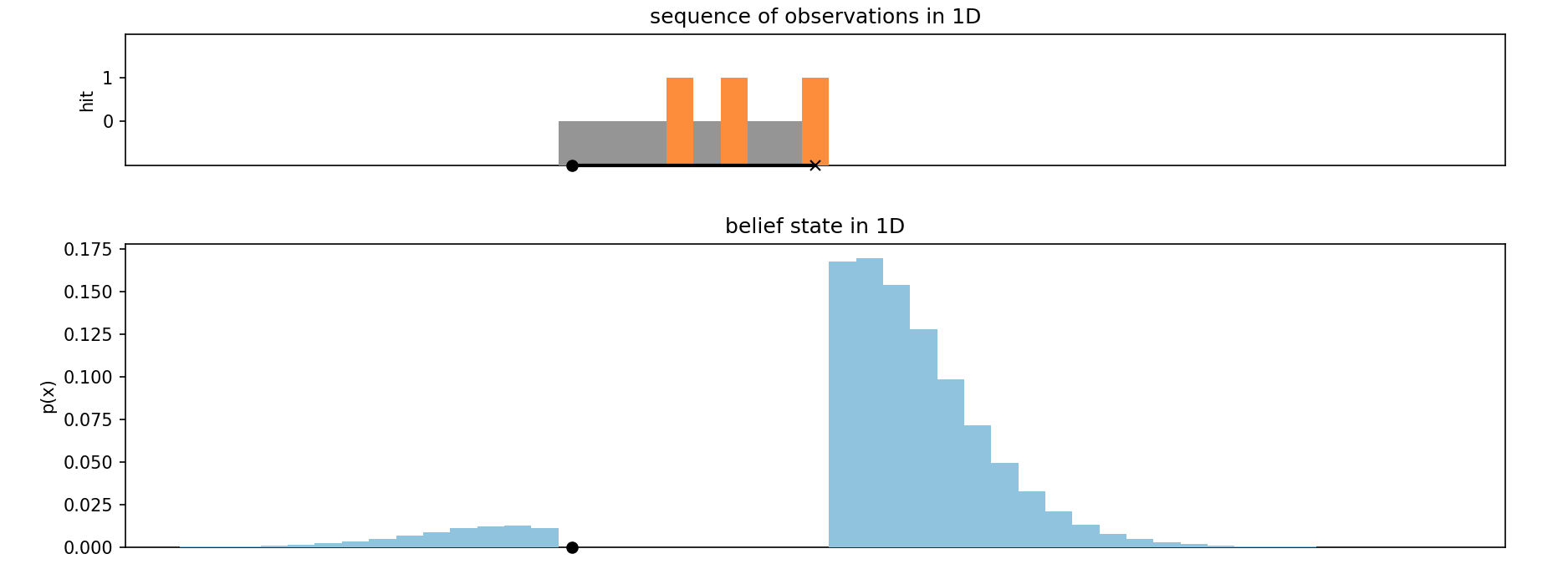} \\[2em]
    \flushleft
    \hspace{1cm} (b) \\
    \centering
    \includegraphics[width=0.99\linewidth]{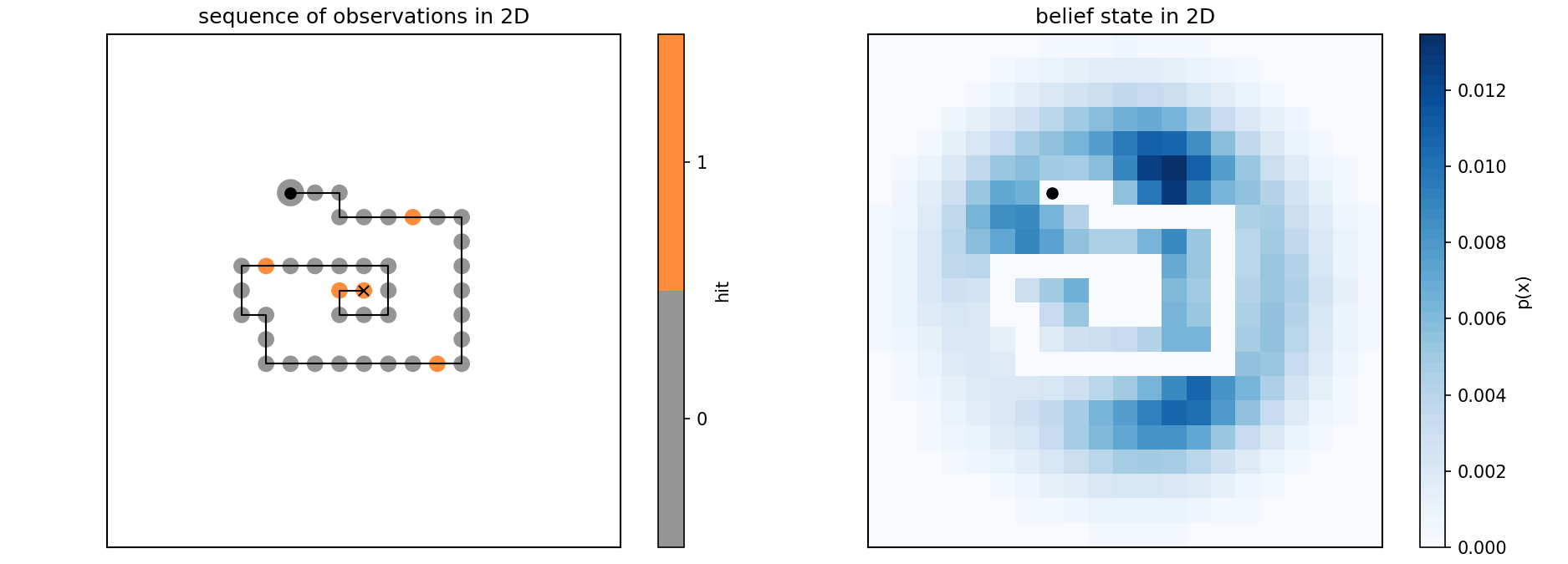} \\[2em]
    \flushleft
    \hspace{1cm} (c) \\
    \centering
    \includegraphics[width=0.99\linewidth]{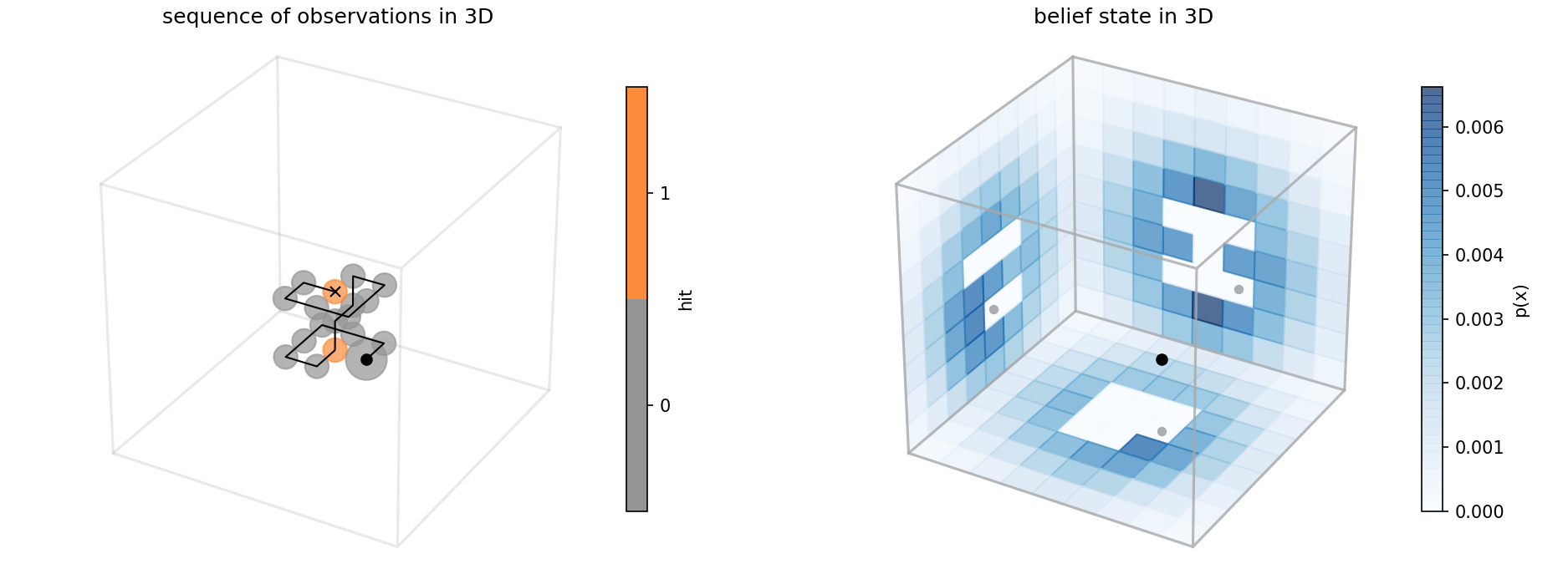}
    \caption{
    Examples of observation sequences and corresponding belief states in (a) 1D, (b) 2D, and (c) 3D. The belief state consists of the agent location (depicted by a black dot) and of a discrete probability distribution over the possible source locations. It accounts for all past observations made by the agent (the agent's initial position is shown by a cross in observation maps). In (c), the source probability distribution is shown in cut-off planes containing the agent.
    \label{fig:illustration_belief_state}
	}
\end{figure*}

The source-tracking problem, illustrated in \cref{fig:illustration_source_tracking_problem}, is a POMDP in which the agent seeks to identify the true location of an imperfectly observed stationary target (a source of odor). The environment is an $n$-dimensional Cartesian grid and the source, invisible to the agent, is located in one of the cells. At each step, the agent moves to a neighbor cell and receives an observation, which provides a noisy measurement of its distance to the source. The search continues until the agent moves to the cell occupied by the source. The agent has a perfect memory and a perfect knowledge of the process that generates observations. How should the agent behave in order to reach the source in the smallest possible number of steps? 

We denote $\v{x}^s$ and $\v{x}^a$ the source and the agent locations ($n$-tuples of integers), respectively. Allowed actions are moves to a neighbor cell along the grid axes, for example in 3D the set of allowed actions is \{'north', 'south', 'east', 'west', 'top', 'bottom'\}. \revision{Staying in the same location is not allowed (unlike in ref. \cite{Vergassola2007}, we made this choice because preliminary tests showed that standing still is almost never beneficial)}. After moving to a cell, either the source is found (event $F$) and the search is over, or the source is not found (complementary event $\bar{F}$) and the agent receives a stochastic sensor measurement $h$ (``hits''), which represents the integer number of odor particles detected by the agent. Hits are received with conditional probability $\Pr(h | \v{x}^a,\v{x}^s)$ (the probabilistic model of detection will be presented in \cref{sec:physical_model_of_detection}). We encompass the presence/absence of the source and the number of hits in a single observation variable $o$. Possible observations are $o \in \{ F, (\bar{F}, 0) , (\bar{F}, 1), (\bar{F}, 2), \dots \}$.

We denote $p(\v{x})$ the discrete probability distribution of the source being in each grid cell, that is $p(\v{x}) = \Pr(\v{x}^s = \v{x})$. 
After each action and observation, $p(\v{x})$ can be updated using Bayesian inference (\cref{sec:Bayes_update}).
The agent has access to its position $\v{x}^a$ and to the distribution $p(\v{x})$. In the POMDP terminology, this defines a belief state $s=[\v{x}^a, p(\v{x})]$. Examples of belief states are shown in \cref {fig:illustration_belief_state}. Finding the source is a special belief state $s^\Omega$ where the source position is known and matches the agent position: $s^\Omega=[\v{x}^a, \delta(\v{x} - \v{x}^a)]$. The agent's behavior is described by a policy, denoted $\pi$, which maps each belief state to an action. For a deterministic policy (as seeked here, since the optimal policy is deterministic), the action chosen is $a=\pi(s)$.

The search proceeds as follows:
\begin{itemize}
    \item Initially
        \begin{itemize} 
         \item The belief state is $s_0=[\v{x}^a_0, p_0(\v{x})]$, where the agent location $\v{x}^a_0$ is at the center of the domain and where the prior distribution of source location $p_0(\v{x})$ is drawn randomly from the set of priors (details are provided in \cref{appsec:initialization_protocol}).
         \item The source location $\v{x}^s$ is drawn randomly according to $p_0(\v{x})$.
        \end{itemize}

    \item At the $t^\text{th}$ step of the search
    \begin{enumerate}
    \item Knowing the current belief state $s_t=[\v{x}^a_t, p_t(\v{x})]$, the agent chooses an action according to some policy $\pi$: $a_{t} = \pi(s_t)$.
    \item The agent moves deterministically to the neighbor cell associated to $a_{t}$. This move is associated to a unit cost. The agent's position is updated to $\v{x}^a_{t+1}$. 
    \item The agent receives an observation $o_{t}$ and the source location distribution is updated, using Bayes' rule, to $p_{t+1}(\v{x}) = \Bayes(p_t(\v{x}), \v{x}^a_{t+1}, o_{t})$ (the Bayes operator will be made explicit in \cref{sec:Bayes_update}).
    \begin{itemize}
        \item If $s_{t+1} = s^\Omega$, that is $\v{x}^a_{t+1} = \v{x}^s$, the search terminates and the agent receives no more costs.
        \item Otherwise, the search continues to step $t+1$.
    \end{itemize}
    \end{enumerate}
\end{itemize}
Each episode (each search) is a sequence like this:
\begin{equation*}
s_0, a_0, o_0, s_1, a_1, o_1, \dots, s_{T-1}, a_{T-1}, o_{T-1}, s^\Omega
\end{equation*}
and the cumulated cost of an episode is equal to the number of steps $T$ to termination (which can be infinite if the source is never found). A step-by-step illustration depicting how a search proceeds is provided in Supplementary Material.

The performance of a policy $\pi$ is measured by $\mathbb{E}_{p_0, \pi} [T]$, the expected number of steps to reach the source, where the expectation is over all possible sequences generated following policy $\pi$, over all possible priors $p_0$, and, implicitely, over all possible source locations which are distributed according to these priors. The optimal policy is the policy that minimizes $\mathbb{E}_{p_0, \pi} [T]$. We expand on the optimal policy in the next section. 

Since the search does not stop until the source is found, any policy that fails to ensure termination with probability one has an infinite cumulated cost.
While it is trivial to show that a policy which guarantees termination exists (any policy that exhaustively searches the grid), the policies we will consider (including infotaxis) do not have such guarantee. Therefore we will evaluate the performance of a policy based on $\Pr(\text{failure})$, the probability of never finding the source, and on $\Mean(T)$, the mean number of steps to reach the source, restricted to episodes where the source is ultimately found. 

Policy evaluation is performed by averaging over a large number of episodes. We use a hybrid Bayesian/Monte-Carlo method which allows faster convergence than traditional Monte-Carlo simulations. Details on our approach are given in \cref{appsec:policy_evaluation} and Supplementary Material. 

\subsection{Optimal policy}
\label{sec:optimal_policy}

Solving the source-tracking problem means finding the optimal policy $\pi^*$ that minimizes the duration of the search
\begin{equation}
 \pi^* = \argmin_\pi \mathbb{E}_{p_0, \pi} [T].
\end{equation}
The optimal policy can, at least formally, be determined from the solution of a recurrence equation known as the Bellman optimality equation as follows.

The optimal value function $v^*(s)$ of a belief state $s$ is defined as the minimum, over all policies, of the expected number of steps remaining to find the source when starting from belief state $s$:
\begin{equation}
 v^*(s) = \min_\pi v^\pi(s) \qquad \text{where} \quad v^\pi(s) = \mathbb{E}_{\pi} [T - t | s_t = s].
\end{equation}

\begin{figure}
    \centering
	\includegraphics[height=10em]{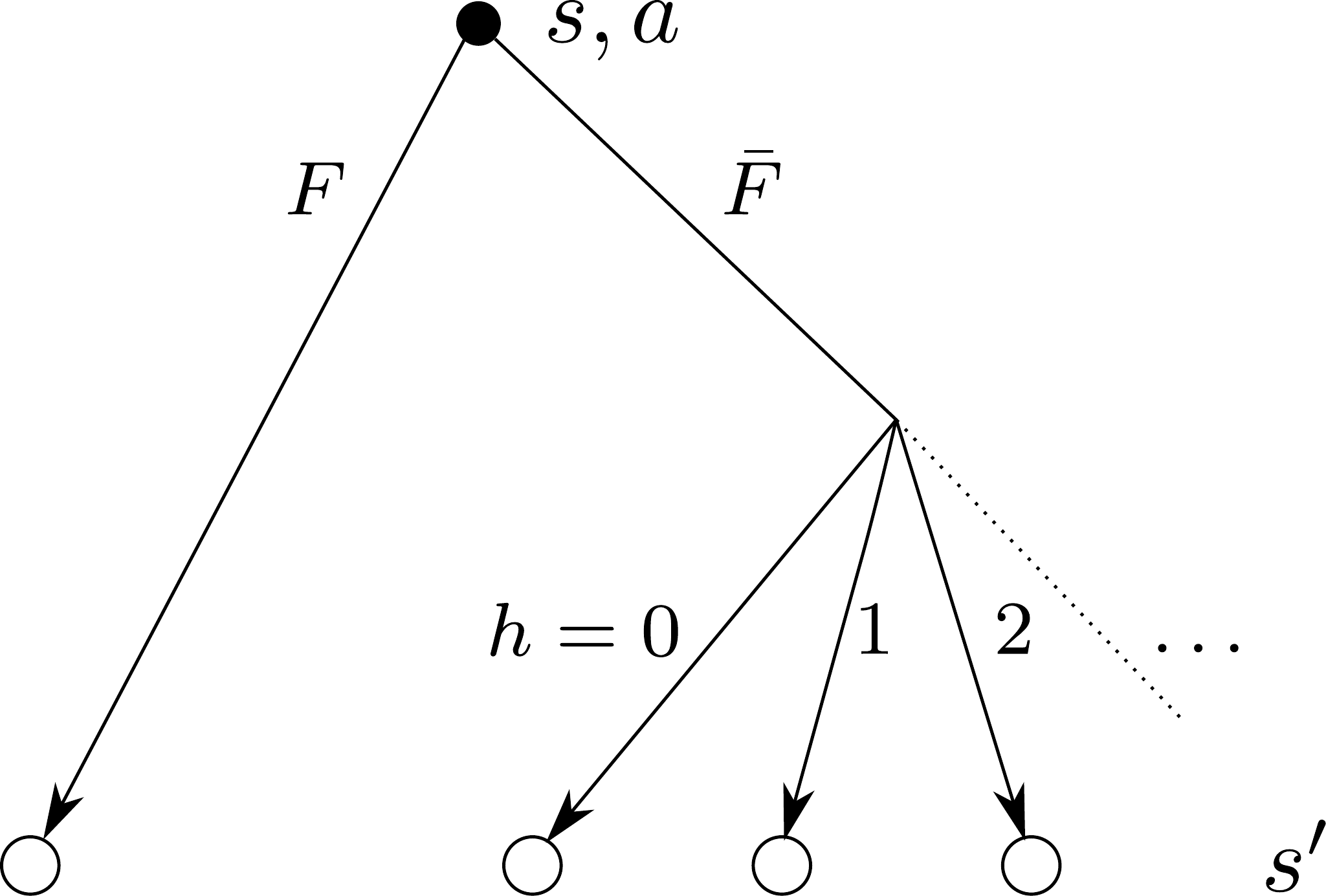}
    \caption{
    Tree of possible successor belief states $s'$ starting from a belief state $s$ and executing action $a$. Transitions from $s$ to $s'$ are determined by the observations: either finding the source ($F$) or not finding the source ($\bar{F}$) and receiving $h$ hits.
	\label{fig:transitions_tree}
	}
\end{figure}

It satisfies the Bellman optimality equation:
\begin{equation}
\label{eq:optimal_Bellman}
    v^*(s) = \min_a \sum_{s'} \Pr(s'|s,a) [1 + v^*(s')]    \qquad  \forall s \neq s^\Omega
\end{equation}
where $\Pr(s'|s,a)$ is the probability of transitioning from belief state $s$ to next belief state $s'$ after executing action $a$, and where $v^*(s^\Omega) = 0$ (by definition of the terminal state $s^\Omega$). Possible transitions from $s$ to $s'$ corresponds to possible observations: either finding the source ($F$) or not finding the source ($\bar{F}$) and receiving $h$ hits, as illustrated in \cref{fig:transitions_tree}.

Given $v^*(s)$, the optimal policy consists in choosing the action that minimizes the expected number of remaining steps $v^*(s')$:
\begin{equation}
\label{eq:optimal_policy_definition}
    \pi^*(s) = \argmin_a \sum_{s'} \Pr(s'|s,a) [1 + v^*(s')].
\end{equation}
When the number of states $s$ is finite (and small enough), \cref{eq:optimal_Bellman} can be solved using dynamic programming \cite{Bellman1957book,Sutton2018book}. In the source-tracking problem however, $s$ contains a discrete probability distribution, of which there are infinitely many. More precisely, there are as many belief states as there are possible histories (sequences of actions and observations, which can be infinitely long). Even if a maximum duration $t_{\text{max}}$ is imposed, the number of possible beliefs grows as $(A \times O)^{t_{\text{max}}}$ where $O$ is the number of possible observations (at least 3) and $A$ is the number of possible actions (twice the number of dimensions). If we consider a 2D domain of $10$ by $10$ cells: $A=4$, $O=3$. With $t_{\text{max}}=100$, the number of possible beliefs is larger than the number of atoms in the observable universe, which prevents from computing the optimal policy exactly. It can, however, be approximated (\cref{sec:RL}). 

A lower bound for $\mathbb{E}_{p_0, \pi^*} [T]$ can be obtained by considering an omniscient agent that has access to the source location. In that case the optimal policy is simply to follow the shortest path to the source, and $T$ is the Manhattan distance $\normone{\v{x}^s - \v{x}^a}$ between the agent and the source.
The lower bound is obtained by taking the expectation of $T$ over the initial source distribution:
\begin{equation}
\label{eq:lower_bound}
 \mathbb{E}_{p_0, \pi^*} [T] \geqslant \mathbb{E}_{p_0} \left[ \sum_{\v{x}} p_0(\v{x}) \normone{\v{x} - \v{x}^a_0} \right].
\end{equation}
Note that this lower bound is not tight since it assumes full observability, however it should be approached as the rate of information received by the agent increases (that is, as the source rate of emission increases).

An upper bound for $\mathbb{E}_{p_0, \pi^*} [T]$ can be obtained by considering an agent without sensors (no information provided by hits) that searches the domain exhaustively following a predetermined trajectory. The optimal trajectory in this scenario is an optimization problem on its own \cite{Baeza-Yates1993}. However a simple yet efficient way to cover space is by ``spiraling'' outward from the starting point (visiting all cells within a Chebyshev distance $d$ of the agent's initial position, then covering all cells at a distance $d+1$, and so on, avoiding as much as possible visiting the same cell twice). In 1D, such spirals are not efficient (the constant back-and-forth motion of the agent would result in a quadratic scaling of the search time with the domain size). An alternative simple trajectory consists in going to one end of the grid and then to the other end. In either case, we call this policy $\pi^{\text{exhaustive}}$ and we can write
\begin{equation}
\label{eq:upper_bound}
 \mathbb{E}_{p_0, \pi^*} [T] \leqslant \mathbb{E}_{p_0, \pi^{\text{exhaustive}}} [T] = \sum_{\v{x}} p_0(\v{x}) r(\v{x})
\end{equation}
where $r(\v{x})$ is the time step at which location $\v{x}$ is visited following the prescribed trajectories (the first visited cell corresponds to $r=1$, the next one to $r=2$, etc.).

\subsection{Probabilistic model of detections}
\label{sec:physical_model_of_detection}

We now need to specify the model used to generate observations (hits), that is, to specify $\Pr(h | \v{x}^a,\v{x}^s)$. This model is based on a physical modeling of dispersion and detection in a turbulent flow as follows.

The source emits detectable odor particles with a finite lifetime which disperse in the ambient turbulent environment and can be detected by the searcher. The detection events, or ``hits'', are distributed according to a Poisson's law
\begin{equation}
 \Pr(h | \mu) = \frac{\mu^h \exp(-\mu)}{h!}
\end{equation}
where $\mu$ is the mean number of hits. It is a function of the Euclidean distance $d=\normtwo{\v{x}^s - \v{x}^a}$ between the agent and the source such that
\begin{equation}
 \Pr(h | \v{x}^a,\v{x}^s) = \Pr(h | \mu(d)) \, \, \text{with} \, d = \normtwo{\v{x}^s - \v{x}^a}.
\end{equation}

The derivation of $\mu(d)$ for an arbitrary number of dimensions $n$ is given in Supplementary Material. 
The resulting expressions are provided below:
\begin{subequations}
 \label{eq:mu_definition}
\begin{align}
 & n=1: && \displaystyle \mu(d) = R \Delta t \frac{\lambda}{\lambda-a} \exp(-d/\lambda) \\
 & n=2: && \displaystyle \mu(d) = R \Delta t \frac{1}{\ln(\lambda/a)} K_{0} (d/\lambda) \\
 & n=3: && \displaystyle \mu(d) = R \Delta t \frac{a}{d} \exp(-d/\lambda) \\
 & n=4: && \displaystyle \mu(d) = R \Delta t \left( \frac{a}{\lambda} \right)^{2} \frac{\lambda}{d} K_{1} (d/\lambda)
\end{align}
and more generally for $n\geqslant 3$
\begin{equation}
 \mu(d) = R \Delta t \left( \frac{a}{\lambda} \right)^{n-2} \left( \frac{\lambda}{d} \right)^{n/2-1} \frac{(n-2)}{\Gamma(n/2)} \frac{K_{n/2-1} (d/\lambda)}{2^{n/2-1}} 
\end{equation}
\end{subequations}
where $a$ is the agent radius, $\lambda$ is the dispersion lengthscale of the particles in the medium, $R$ is the source emission rate, $\Delta t$ is the duration of a sensor measurement, $\Gamma$ is the gamma function, and $K_{\nu}$ is the modified Bessel function of the second kind of order $\nu$.
Note that we recover the expressions provided in ref.~\cite{Vergassola2007} for $n=2$ and $n=3$.

\subsection{Update by Bayesian inference}
\label{sec:Bayes_update}

Each observation provides some information about the source location, which can be accounted for using Bayesian inference. In the case of a sequential process such as ours, Bayes' rule can be applied after each observation to maintain an up-to-date belief $p(\v{x})$ which encompasses all information gathered so far. The update after observing $o_t$ in $\v{x}^a_{t+1}$ reads
\begin{equation}
    p_{t+1}(\v{x}) = \Bayes(p_t(\v{x}), \v{x}^a_{t+1}, o_{t})
\end{equation}
where $\Bayes(p(\v{x}), \v{x}^a, o)$ is the operator that maps the prior $p_t$ to the posterior $p_{t+1}$ through Bayes' rule
\begin{equation}
\label{eq:Bayes_update_generic}
 \Bayes(p(\v{x}), \v{x}^a, o) = \frac{\Pr(o | \v{x}^a,\v{x}) p(\v{x})}{\sum_{\v{x}'} \Pr(o | \v{x}^a,\v{x}') p(\v{x}')}
\end{equation}
and where $\Pr(o | \v{x}^a,\v{x})$ is called the evidence in Bayesian terminology.

Let us now go through the update rule for each observation.
If $o=F$, the source has been found in $\v{x}^a$, and the posterior distribution is simply a Dirac distribution
\begin{equation}
 \Bayes(p(\v{x}), \v{x}^a, F) = \delta(\v{x} - \v{x}^a).
\end{equation}
Otherwise, $o=(\bar{F}, h)$, meaning that the source has not been found and that $h$ hits were perceived. 
The posterior distribution after not finding the source is a simple renormalization
\begin{equation}
\label{eq:Bayes_update_not_found}
 \Bayes(p(\v{x}), \v{x}^a, \bar{F}) = 
 \begin{dcases}
    0 & \text{if $\v{x} = \v{x}^a$,} \\ 
    \frac{p(\v{x})}{\sum_{\v{x}' \neq \v{x}^a} p(\v{x}')} & \text{otherwise.}
\end{dcases}
\end{equation}
The posterior after a hit $h$ is
\begin{equation}
\label{eq:Bayes_update_hit}
    \Bayes(p(\v{x}), \v{x}^a, h) = \frac{\Pr(h | \v{x}^a,\v{x}) \, p(\v{x})}{\sum_{\v{x}'} \Pr(h | \v{x}^a,\v{x}') \, p(\v{x}')}.  
\end{equation}
The full update after observing $o=(\bar{F}, h)$ is therefore given by the successive application of each partial update
\begin{equation}
 \Bayes(p(\v{x}), \v{x}^a, o) = \Bayes ( \Bayes(p(\v{x}), \v{x}^a, \bar{F}) , \v{x}^a, h ).
 \end{equation}
A step-by-step illustration of Bayesian updates performed during a search is provided in Supplementary Material.

\subsection{Parameters of the problem}

The source-tracking problem is parameterized by:
\begin{enumerate}
 \item the space resolution $\Delta x$ (the size of the agent step, also the linear size of a grid cell);
 \item the time resolution $\Delta t$ (how often does the agent make a decision, also the integration time for the sensors); or alternatively the agent speed $v$ and then $\Delta t = \Delta x / v$;
 \item the probabilistic law for hits encounters, itself parameterized by $(a,\lambda,R,\Delta t)$, where $a$ is the agent radius, $\lambda$ is the characteristic lengthscale of dispersion, $R$ is the source emission rate;
 \item the initial conditions (prior source distribution and agent's position);
 \item the grid size $N$ (linear size of the domain).
\end{enumerate}

The number of parameters can be greatly reduced by assuming that the search is initialized by a (nonzero) hit. The advantages are twofold: (i) the definition of the start is not arbitrary but instead corresponds to the moment when the agent is informed that there is source (as opposed to nothing) in the neighborhood and where tracking it down becomes meaningful, and (ii) the grid size can be chosen large enough such that the domain boundaries play virtually no role in the search (this is because the source probability distribution after a hit decays exponentially with the distance to the agent). Note that mimicking an open world is highly desirable since the physical problem of dispersion is modeled for an infinite space and any finite-size effect in this context would be physically irrelevant. The details of this initialization procedure are provided in \cref{appsec:initialization_protocol}. We will also assume that the size of the agent step is equal to its body diameter $\Delta x = 2 a$. 

With our initialization protocol, the source-tracking problem involves only four parameters which govern the physics of propagation and detection ($\Delta x$, $\lambda$, $R$, $\Delta t$). From those we can construct two independent dimensionless numbers, we chose
\begin{equation}
\label{eq:def_dimensionless_parameters}
    \mathcal{L} = \frac{\lambda}{\Delta x} \qquad \text{and} \qquad \mathcal{I} = R \Delta t
\end{equation}
which characterize the problem size and the source intensity, respectively. 

Together with the problem dimensionality (1D, 2D, etc.), $\mathcal{L}$ and $\mathcal{I}$ are the relevant physical parameters of the source-tracking POMDP.

\subsection{Source-tracking is hard}
\label{sec:naive_policies}

To get a sense of the difficulty of the source-tracking problem, it is worth reporting the performance of simple (and somewhat naive) heuristic policies. Examples of such policies are the usual greedy policy (which maximizes the probability of finding the source at the next step), and policies that would be optimal in the absence of uncertainty (such as the most likely state and the voting policies, both proposed by Cassandra et al.~\cite{Cassandra1996}, and a policy that minimizes the expected distance to the source at the next step).

The description and the performance of these four policies are provided in Supplementary Material. Reliability is very poor for all of them, with $\Pr(\text{failure})$ typically larger than 1 \% or even 10 \%. The greedy policy, which is locally optimal, leads the agent astray when few detections occur. Policies that would be optimal in the absence of uncertainty are plagued by the emergence of loops in the agent's trajectories. These loops arise because of the high degree of symmetry in $p(\v{x})$, due to the absence of directionality in the information provided by hits. Videos illustrating the behavior of these different policies are provided in Supplementary Material.

\section{The infotaxis policy}
\label{sec:infotaxis}

Infotaxis \cite{Vergassola2007} is a heuristic solution to the source-tracking problem which greedily minimizes uncertainty on the source location: at each step, an infotactic agent chooses the action that maximizes the expected information gain. 
In this section we describe the infotaxis policy and evaluate its performance. 

\subsection{Description of infotaxis}

The (Shannon) entropy of a belief state $s=[\v{x}^a, p(\v{x})]$ is defined by
\begin{equation}
    \label{eq:definition_entropy}
    H(s) = - \sum_{\v{x}} p(\v{x}) \log_2 p(\v{x})
\end{equation}
and is a measure of how uncertain is the source location (we use the logarithm with base 2, as is standard in information theory where $H$ is measured in bits of information). In particular, the entropy of a Dirac distribution is zero, so $H(s^\Omega)=0$. Note that $H$ is independent of the agent's position $\v{x}^a$.

The expected entropy upon taking action $a$ in belief state $s$, denoted $H(s|a)$, is the expected entropy of successor belief states $s'$:
\begin{equation}
    \label{eq:H_s_given_a}
    H(s|a) = \sum_{s'} \Pr(s'|s,a) H(s')
\end{equation}
where $\Pr(s'|s,a)$ is the probability of transitioning from belief state $s$ to next belief state $s'$ after executing action $a$ (cf. \cref{fig:transitions_tree}). 
An explicit example showing how $H(s|a)$ is calculated in practice is provided in Supplementary Material.

The information gain associated with action $a$ in belief state $s$ is then given by
\begin{equation}
    \label{eq:def_G}
    G(s,a) = H(s) - H(s | a)
\end{equation}
Note that the information gain is also, by definition \cite{MacKay2005book}, the mutual information between the current belief state $s$ and the possible observations after action $a$. 
The infotaxis policy, denoted $\pi^{\text{infotaxis}}$, consists in choosing the action that maximizes the information gain (or equivalently, the action that minimizes the expected entropy) at the next step, that is,
\begin{equation}
 \label{eq:pi_infotaxis}
 \pi^{\text{infotaxis}}(s) = \argmax_a G(s,a) = \argmin_a H(s | a).
\end{equation}

Combining \cref{eq:H_s_given_a} with \cref{eq:pi_infotaxis}, the infotaxis policy can be written as 
\begin{equation}
 \pi^{\text{infotaxis}}(s) = \argmin_a \sum_{s'} \Pr(s'|s,a) H(s')
\end{equation}
which makes clear that it shares the same structure as the optimal policy, given by \cref{eq:optimal_policy_definition} and which can be rewritten as
\begin{equation}
    \pi^*(s) = 1 + \argmin_a \sum_{s'} \Pr(s'|s,a) v^*(s').
\end{equation}
If $v^*(s)$, the expected time to find the source starting from belief state $s$ with the optimal policy, was a linear function of $H(s)$, then infotaxis would be optimal. A trivial counter-example shows that this is not the case: when the source location is known, $v^*$ is equal to the Manhattan distance to the source whereas $H=0$ for all source locations. 

\subsection{Performance of infotaxis}

\begin{figure*}
    \flushleft
    \hspace{2.5cm} (a) \hspace{7cm} (b) \\
    \centering
    \includegraphics[width=0.33\linewidth]{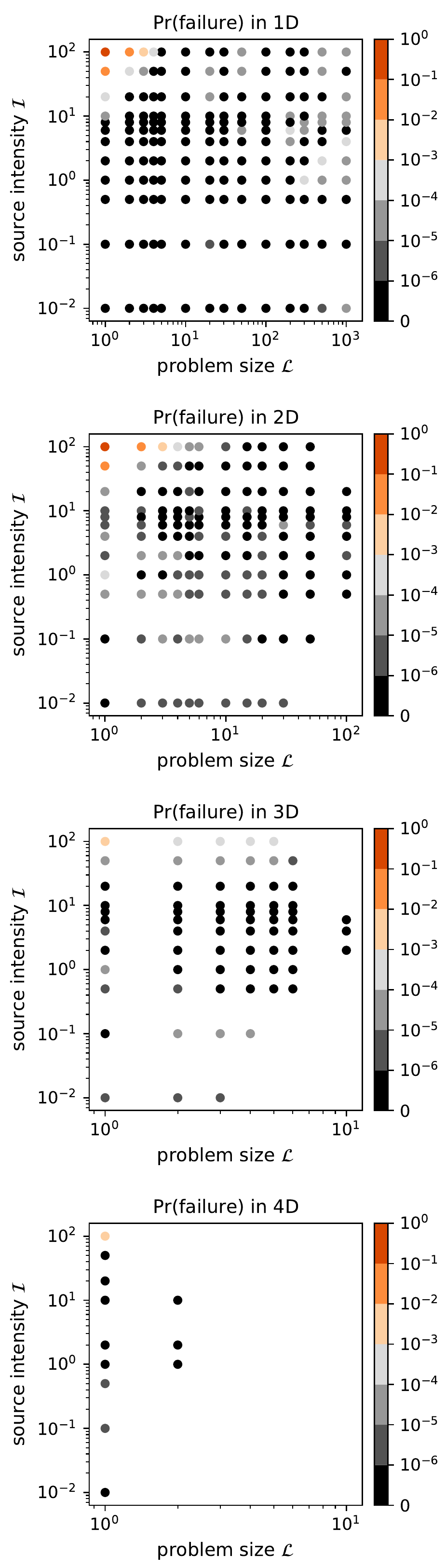}
    \hspace{1.5cm}
    \includegraphics[width=0.33\linewidth]{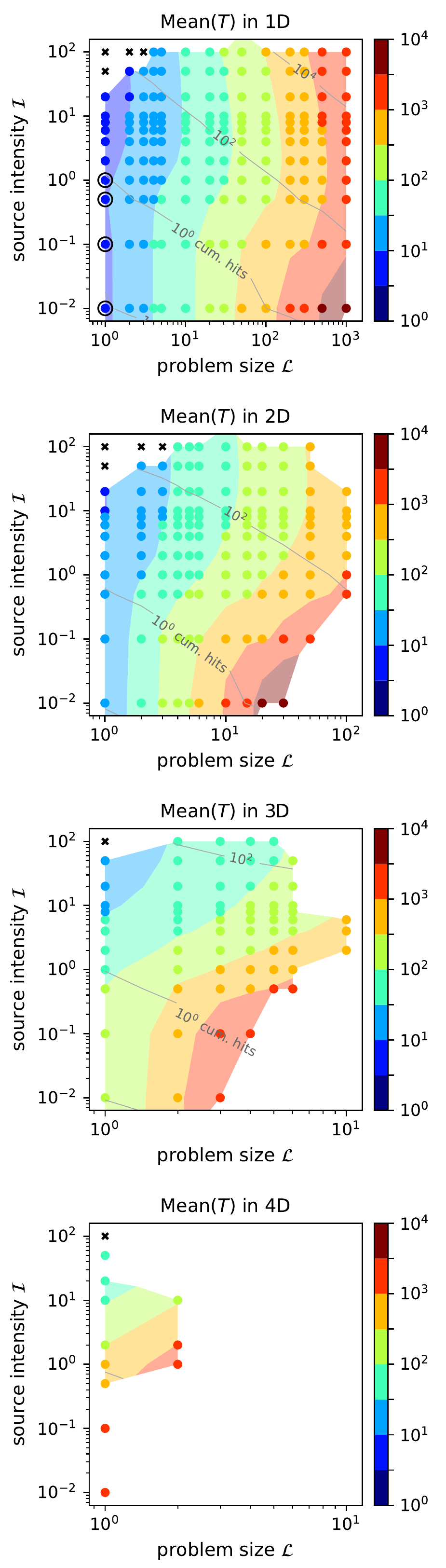}
	\caption{
	Performance of infotaxis for source tracking in 1D, 2D, 3D and 4D (rows), for a wide range of problem sizes $\mathcal{L}$ and source intensities $\mathcal{I}$: (a) probability of never finding the source and (b) mean number of steps to find the source. In (b), the black crosses depict cases where $\Pr(\text{failure}) > 10^{-3}$, the grey contour lines indicate the mean cumulated number of hits gathered along the search (the detailed data is provided in Supplementary Material), and the outer circles depict cases where $\Mean(T)$ is larger than the upper bound we obtained for the optimal policy.
	\label{fig:infotaxis_performance}
	}
\end{figure*}

\begin{figure*}
    \flushleft
    \hspace{2.5cm} (a) \hspace{7cm} (b) \\
    \centering
    \includegraphics[width=0.33\linewidth]{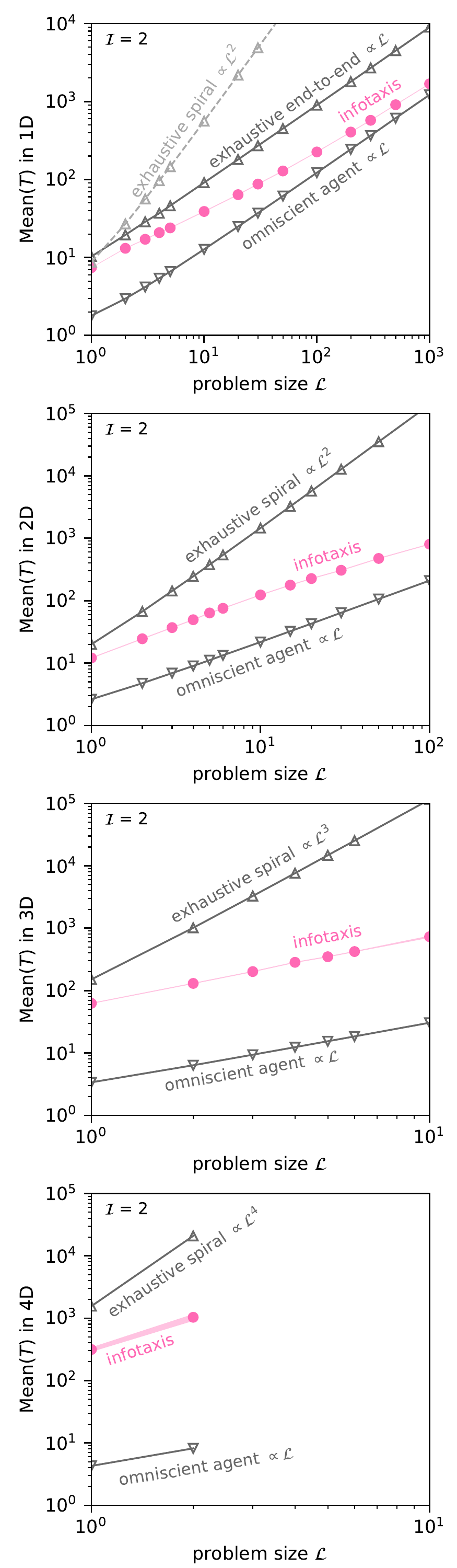}
    \hspace{1.5cm}
    \includegraphics[width=0.33\linewidth]{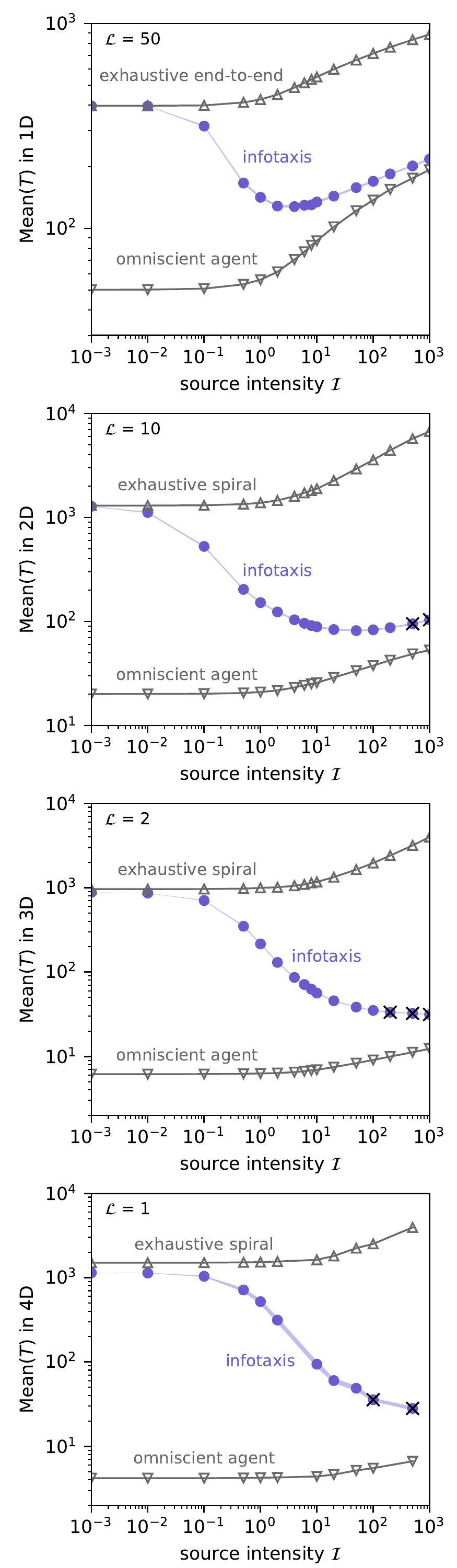}
	\caption{
	Mean number of steps to find the source with infotaxis in 1D, 2D, 3D and 4D (rows) as a function of (a) the problem size $\mathcal{L}$ and (b) the source intensity $\mathcal{I}$. Infotaxis is depicted by dots (crossed dots indicate $\Pr(\text{failure}) > 10^{-3}$), while downward (resp. upward) triangles indicate the lower (resp. upper) bound for the optimal policy computed according to \cref{eq:lower_bound} (resp. \cref{eq:upper_bound}). The lower bound is strict and assumes an omniscient agent. The upper bound corresponds to an exhaustive search following an outward spiral, except in 1D where an ``end-to-end'' trajectory is better (unless the domain is very small). The (thin) shaded area between infotaxis dots show 95 \% confidence intervals.
	\label{fig:infotaxis_bounds}
	}
\end{figure*}

\begin{figure*}
	\centering
    \includegraphics[width=0.9\linewidth]{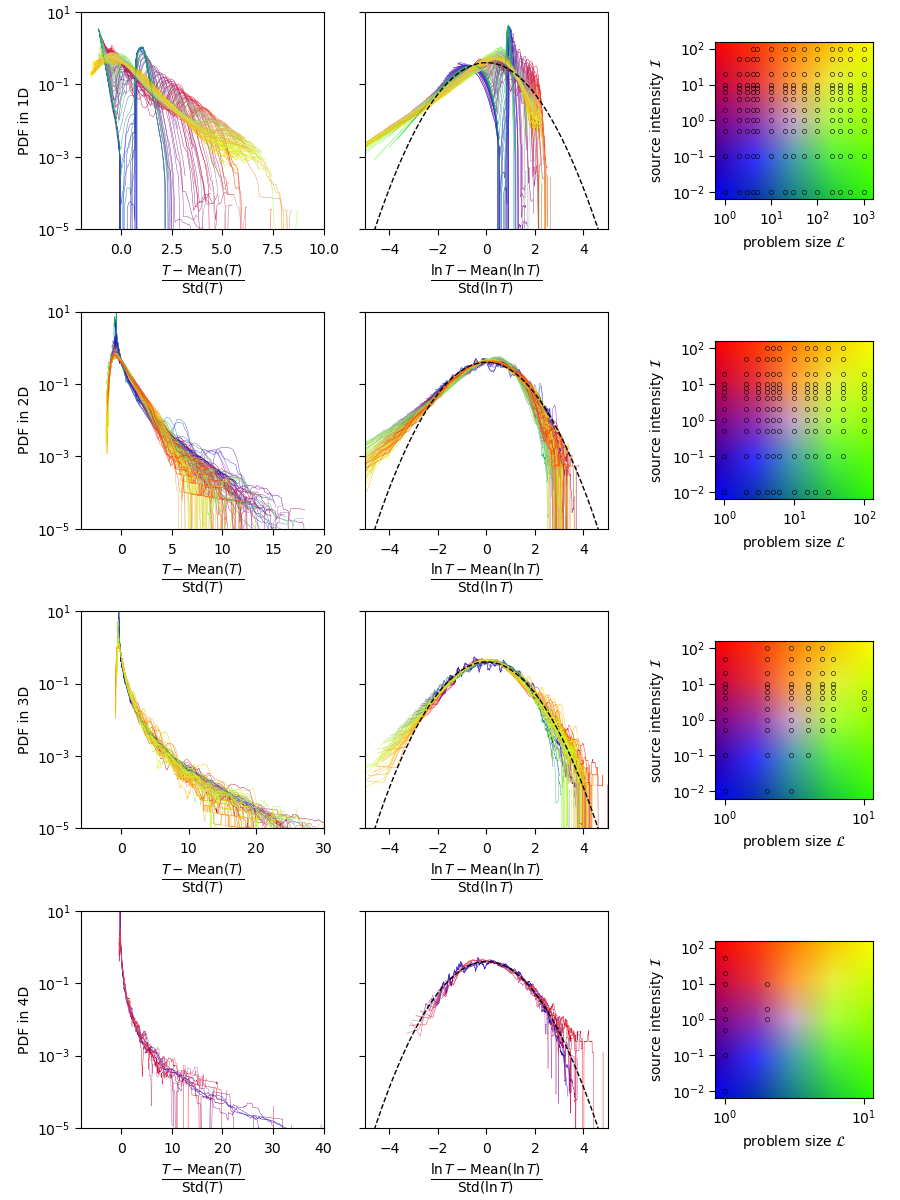}
	\caption{
	Distributions of arrival times with infotaxis for source tracking in 1D, 2D, 3D and 4D (rows) for a wide range of problem sizes $\mathcal{L}$ and source intensities $\mathcal{I}$ (color-coded according to the map on the right). The same data is plotted in both columns using different scales to evidence the distribution features. In the left column, exponential decay would appear as linear. In the right column, a standardized lognormal distribution would appear as a parabola (dashed curve), which corresponds to a standard Gaussian in the log-lin space.
	\label{fig:infotaxis_pdfs}
	}
\end{figure*}

We now provide an assessment of the infotaxis policy. The performance of infotaxis has been evaluated in $n=\{1,2,3\}$ dimensions and for a wide range of problem sizes $\mathcal{L}$ and source intensities $\mathcal{I}$, defined by \cref{eq:def_dimensionless_parameters}. A few additional simulations have also been performed in 4D. The results are summarized in \cref{fig:infotaxis_performance}. Missing data correspond to cases we cannot simulate, either because they require too much memory or because they are too expensive computationally. Videos showing infotactic searches are provided in Supplementary Material.

The first quantity of interest is the probability of failing to find the source, $\Pr(\text{failure})$, which is plotted in the $(\mathcal{L},\mathcal{I})$ plane in \cref{fig:infotaxis_performance}(a). In our simulations, cases with $\Pr(\text{failure}) \leqslant 10^{-6}$ have negligible probability of failure within our numerical accuracy (black dots). Conversely, cases with $\Pr(\text{failure}) > 10^{-6}$ (grey to orange dots) exhibit a measurable probability of never finding the source, which is due to the agent entering an infinite loop at some point during the search (which then never terminates). It is clear from \cref{fig:infotaxis_performance}(a) that infotaxis does not always guarantee termination, however $\Pr(\text{failure}) < 10^{-3}$ (black to grey shades) for almost all cases considered. The exceptions (orange shades) correspond to situations with small $\mathcal{L}$ and large $\mathcal{I}$: these are searches where hits provide so much information that the source location can be identified exactly from a distance. In this situation, the entropy of the belief state is zero and cannot be reduced any further: the infotactic agent is then ``lost''. This marginal issue could be easily fixed by requiring the agent to go to the source if its location is perfectly known.

The second quantity of interest is the mean number of steps to reach the source, provided that the source is found with sufficiently high probability (we set this threshold arbitrarily to $0.999$). $\Mean(T)$ is plotted in the $(\mathcal{L},\mathcal{I})$ plane using a colormap in \cref{fig:infotaxis_performance}(b), where we also show grey contours for the mean cumulated number of hits (excluding the initial hit). A more quantitative representation of $\Mean(T)$ is presented in \cref{fig:infotaxis_bounds}, where we also show the theoretical lower and upper bounds we obtained for the optimal policy (\cref{eq:lower_bound,eq:upper_bound}, respectively). 

The upper bound was obtained by considering a search where the agent explores the space exhaustively by following n-dimensional spiralling trajectories, or, in 1D, by going to one end and then to the other end of the domain (unless the one-dimensional spiral is better, which is the case only for very small domains). This upper bound therefore scales as $\mathcal{L}^n$ (with $n$ the dimensionality), as illustrated by \cref{fig:infotaxis_bounds}(a).
Cases where $\Mean(T)$ is larger than this upper bound are highlighted in \cref{fig:infotaxis_performance}(b) by outer circles: the mean search time achieved by infotaxis is never above the upper bound for the optimal policy, except in 1D for small $\mathcal{L}$ and $\mathcal{I}$ (a rather marginal configuration).

Besides, it has been shown that in the absence of detections (succession of zero hits), infotactic trajectories are Archimedean spirals in 2D \cite{Vergassola2007} and an approximate generalization of those in 3D \cite{Masson2009,Barbieri2011}. We recover such trajectories, they are shown in \cref{appsec:zerohit}. However, in the 1D case, we find that infotaxis does not generate one-dimensional spirals but instead ``end-to-end'' trajectories in the absence of hits. These trajectories are essentially the ones we assumed to derive our upper bounds: the latter are therefore a very good estimate of $\Mean(T)$ in the limit of no information $\mathcal{I} \rightarrow 0$, as can be seen from \cref{fig:infotaxis_bounds}(b).

The lower bound was obtained by considering an agent which knows the source location, and hence scales linearly in $\mathcal{L}$. We expect the optimal policy to approach this bound when $\mathcal{I}$ becomes large (though this bound is not tight). From \cref{fig:infotaxis_bounds}(b), this is clearly the case for the infotaxis policy. Note that at high $\mathcal{I}$, infotactic agents tend to get stuck (cf. \cref{fig:infotaxis_performance}(a)), which limits the maximal value of $\mathcal{I}$ we can reach. 

Even though $\Mean(T)$ is the quantity to minimize in the source-tracking problem, a strategy exhibiting huge fluctuations in search times may be too hazardous to be used in practice. We found that the standard deviation of the arrival times distribution, $\Std(T)$, is of the same order of $\Mean(T)$ with very little variations across the entire parameter range we explored: $\Std(T) / \Mean(T) \in [0.6, 1.1]$ in 1D, $[0.7, 1.9]$ in 2D, $[1.0, 2.7]$ in 3D, and around 3 in 4D (though the parameter range is limited in this case). The detailed data is provided in Supplementary Material. 

The distributions of arrival times for all considered parameters are presented in \cref{fig:infotaxis_pdfs}. Note that, compared to prior work, our methodology allowed us to properly sample the tail of the distributions up to $O(10)$ standard deviations. Our data (left column of \cref{fig:infotaxis_pdfs}) establish that the tail is subexponential, unlike what was previously believed \cite{Vergassola2007,Masson2009,Ristic2016,Eggels2017}. By plotting the same data using a log-log scale, it can be shown that the tail decays faster than any power law. We found that the arrival time distributions approach log-normal distributions as the number of space dimensions increases. The log-normal probability density function reads
\begin{equation}
 f(T) = \frac{1}{\sigma \tau \sqrt{2 \upi}} \exp \left( - \frac{(\ln T - \mu)^2}{2 \sigma^2} \right) 
 \label{eq:lognormal_def}
\end{equation}
where $\mu = \Mean(\ln T)$ and $\sigma = \Std(\ln T)$. 
The distribution of $\tau=(\ln T - \mu)/\sigma$ is plotted in \cref{fig:infotaxis_pdfs} (central column). If $T$ was perfectly log-normally distributed, $\tau$ would follow a standard normal distribution (which is shown by the dashed curve). 
Log-normal distributions generally result from multiplicative random processes, by application of the central limit theorem in the log domain. Yet, log-normal distributions often appear as a good empirical fit for lifetime or reliability analyses, or more generally for distributions of a positive variable.

To summarize, we computed the statistics of infotactic searches in 1D, 2D, 3D and 4D, for a wide range of dimensionless problem sizes $\mathcal{L}$ and cue emission intensities $\mathcal{I}$. Our data show that infotaxis is \emph{reliable} (the probability of failure is less than $10^{-3}$), \emph{efficient} (the mean search time scales as expected for the optimal policy), and \emph{safe} (the tail of the distribution of search times decays faster than any power law, though subexponentially). 

\subsection{N-step infotaxis}

\begin{figure*}
    \centering
    \includegraphics[width=0.99\linewidth]{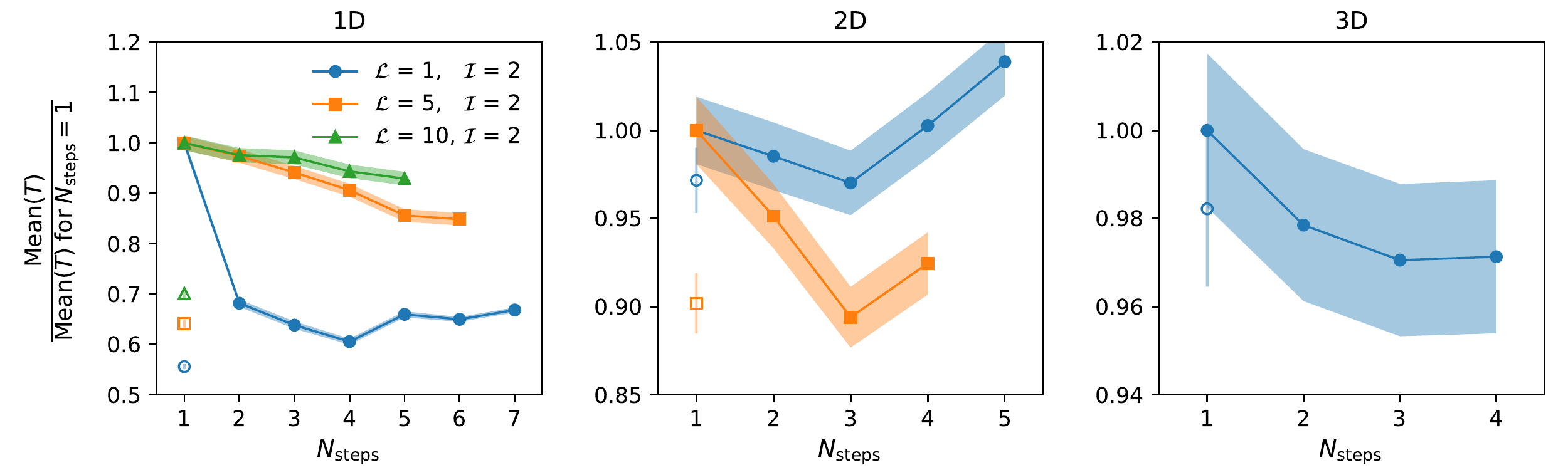}
    \caption{
	Performance of N-step infotaxis in 1D, 2D and 3D (columns): mean number of steps to find the source as a function of the number of anticipated steps $N_{\text{steps}}$, normalized by its value for 1-step infotaxis (filled symbols). The shaded areas represent 95\% confidence intervals. For comparison, we also show the performance of our ``space-aware infotaxis'' with open symbols (cf. \cref{sec:space_aware_infotaxis}).
	\label{fig:infotaxis_vs_Nfuture}
	}
\end{figure*}

Infotaxis is based on a one-step anticipation of possible outcomes of each action. In this section we evaluate N-step infotaxis, the generalization of infotaxis to an anticipation over an arbitrary number of steps.

N-step infotaxis considers all possible outcomes of all possible sequences of $N_{\text{steps}}$ actions, and maximizes the cumulated information gain over those steps. The detailed algorithm, which relies on an exhaustive tree search, is given in Supplementary Material. N-step infotaxis is of very limited practictal interest, because its cost is exponential in $N_{\text{steps}}$. Yet it is fundamentally interesting to determine whether a significant increase of computational power leads to valuable improvements of the search. 

The performance of N-steps infotaxis as a function of the number of anticipated steps is reported in \cref{fig:infotaxis_vs_Nfuture} for searches in 1D, 2D and 3D, for $\mathcal{L}=1$, $5$, or $10$ and $\mathcal{I}=2$. Results are quite variable depending on the case considered. In 1D, improvements are significant for the smallest domain (40 \% reduction in the mean search time for $\mathcal{L}=1$) while more modest for larger domains (15 and 7 \% for $\mathcal{L}=5$ and $10$, respectively, though a plateau is not reached yet). In 2D, the search time is reduced by 3 \% for the smallest domain ($\mathcal{L}=1$), and by 11 \% for a larger one ($\mathcal{L}=5$). The 2D case with $\mathcal{L}=1$ also shows an example where N-step infotaxis ($N_{\text{steps}} = 5$) performs worse than infotaxis. In 3D, improvements plateau at 3 \%, though we tested only the smallest domain size. 

Overall, the search duration first decreases as $N_{\text{steps}}$ increases, and then either reaches a plateau or increases again as $N_{\text{steps}}$ is increased further. This non-monotonic behaviour is consistent with prior experimental results in 2D \cite{Lochmatter2010thesis}, and is another evidence of the absence of a linear correlation between the entropy and the remaining time to find the source. Importantly this minimum (or this plateau) does not generally correspond to the optimal performance: it is clear in 1D that our own heuristic, ``space-aware infotaxis'' (depicted by open symbols, and which will be presented in the next section) performs better.

\section{A better heuristic: space-aware infotaxis}
\label{sec:space_aware_infotaxis}

Infotaxis is inherently risk-averse, which explains why it is so reliable. But this behavior is not optimal: it is indeed known that shifting infotaxis toward more exploitation using a tunable parameter can reduce search duration \cite{Masson2009,Masson2013}. This is however not satisfactory, since the value of this parameter has to be optimized for each set of $(\mathcal{L}, \mathcal{I})$. Here we present a parameter-free policy that achieves better performance than infotaxis.

\subsection{Description of space-aware infotaxis}

Entropy is a quantity that does not contain any information about the source location relative to agent. To make infotaxis more exploitative, we will build a policy based on entropy $H$, which measures uncertainty, and on a spatial metric, denoted $D$, which quantifies the distance between the agent and the source. 

The quantity we want to minimize should naturally balance $H$ and $D$, and be related to the time remaining to find the source. We chose the following expression
\begin{equation}
    J(s) = \log_2 \left( D(s) + 2^{H(s)-1} - \frac{1}{2} \right)
\end{equation}
and $J(s^\Omega)=0$. It is a function of $H(s)$, the entropy of belief state $s$, defined in \cref{eq:definition_entropy}, and of $D(s)$, the mean Manhattan distance between the agent and the source
\begin{equation}
    \label{eq:def_D_Manhattan}
    D(s) = \sum_{\v{x}} p(\v{x}) \normone{\v{x} - \v{x}^a}
\end{equation}
where $p(\v{x})$ is the probability of the source being in $\v{x}$ and $\v{x}^a$ is the agent location.
This seemingly \textit{ad hoc} expression of $J$ has been constructed from the following hypotheses. 

Consider a distribution $p(\v{x})$ with entropy $H$. Assuming $p(\v{x})$ is uniform, this corresponds to an effective number of cells $N_{\text{eff}} = 2^H$. The expected time to find a uniformly distributed source by visiting exhaustively $N_{\text{eff}}$ cells is
\begin{equation}
    \label{eq:space_aware_infotaxis_effect_of_entropy}
    \mathbb{E}[T] = \sum_{i=0}^{N_{\text{eff}}-1} \frac{i}{N_{\text{eff}}} = \frac{N_{\text{eff}} - 1}{2} = 2^{H-1} - \frac{1}{2}.
\end{equation}
This expression does not take into account any spatial constraint and is derived assuming the agent can jump from cell to cell for a unit cost. 
On the other hand, if the source location is known, the minimal time needed to reach a source located at a distance $D$ is $D$.
Assuming the effects of distance and of uncertainty combine linearly, this gives the expression $D + 2^{H-1} - \frac{1}{2}$ as an estimate of the number of time steps remaining to find the source.

An additional consideration is the mathematical properties that we would like $J$ to satisfy. For any belief-MDP with a cost to minimize, the optimal value function $v^*(s)$ of a belief state $s$, defined by \cref{eq:optimal_Bellman}, is a concave function of that belief state \cite{Sondik1971thesis,Smallwood1973}. Since $J$ is designed to be an approximation of the optimal value function $v^*$, we applied a logarithm function to ensure that $J$ is a concave function of $p(\v{x})$. The particular choice of the logarithm is motivated by the fact that infotaxis is recovered if one enforces $D=0$. Note that the entropy $H$ is indeed a concave function of $p(\v{x})$.

We now introduce ``space-aware infotaxis'' (SAI), the policy that minimizes the expected $J$ at the next step. 
It reads
\begin{equation}
 \pi^{\text{SAI}}(s) = \argmin_a J(s | a)
\end{equation}
where $J(s | a)$ is the expected $J$ upon executing action $a$ in belief state $s$, which is given by
\begin{equation}
 J(s | a) = \sum_{s'} \Pr(s'|s,a) J(s')
\end{equation}
where the sum is taken over all successor belief states $s'$ (cf. \cref{fig:transitions_tree}). Note that SAI looks one step ahead into possible futures, the same way infotaxis does, therefore its computational cost is essentially the same as infotaxis (the only extra computation required is that of $D$). 

\begin{figure*}
    \flushleft
    \hspace{2.2cm} (a) \hspace{7.3cm} (b) \\
    \centering
    \includegraphics[width=0.35\linewidth]{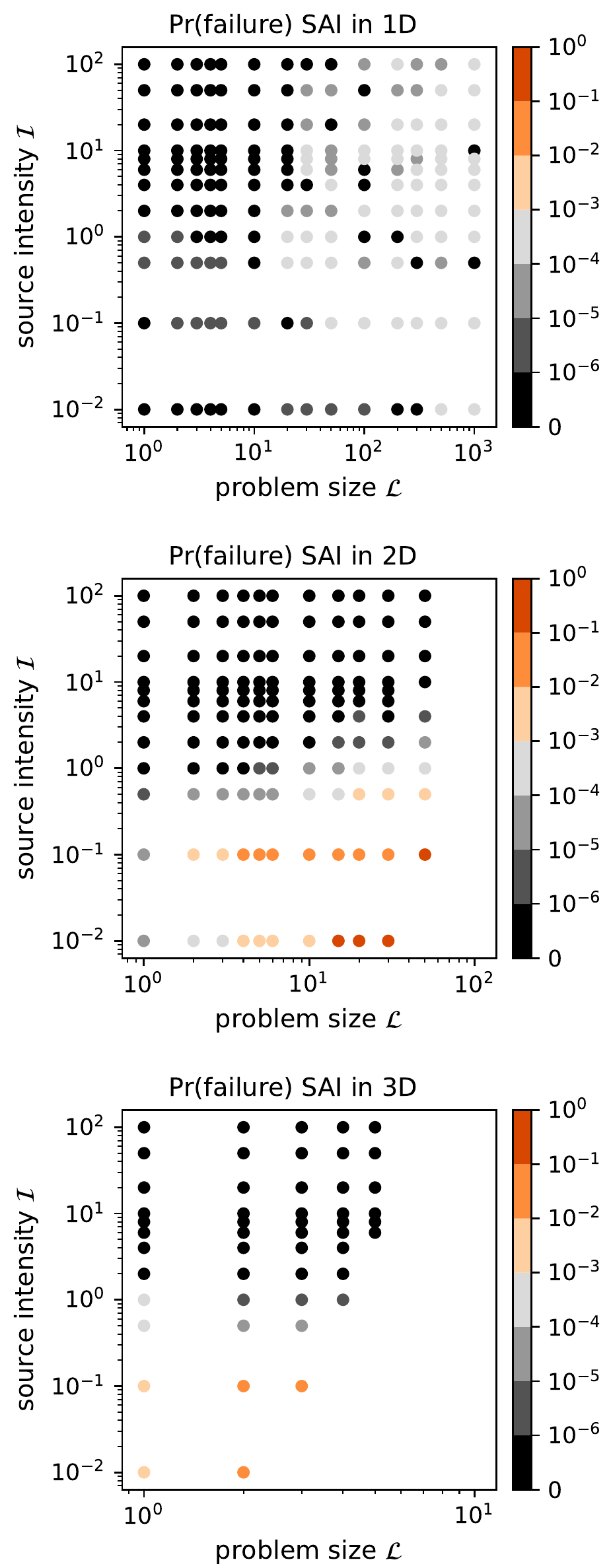}
    \hspace{1.5cm}
    \includegraphics[width=0.35\linewidth]{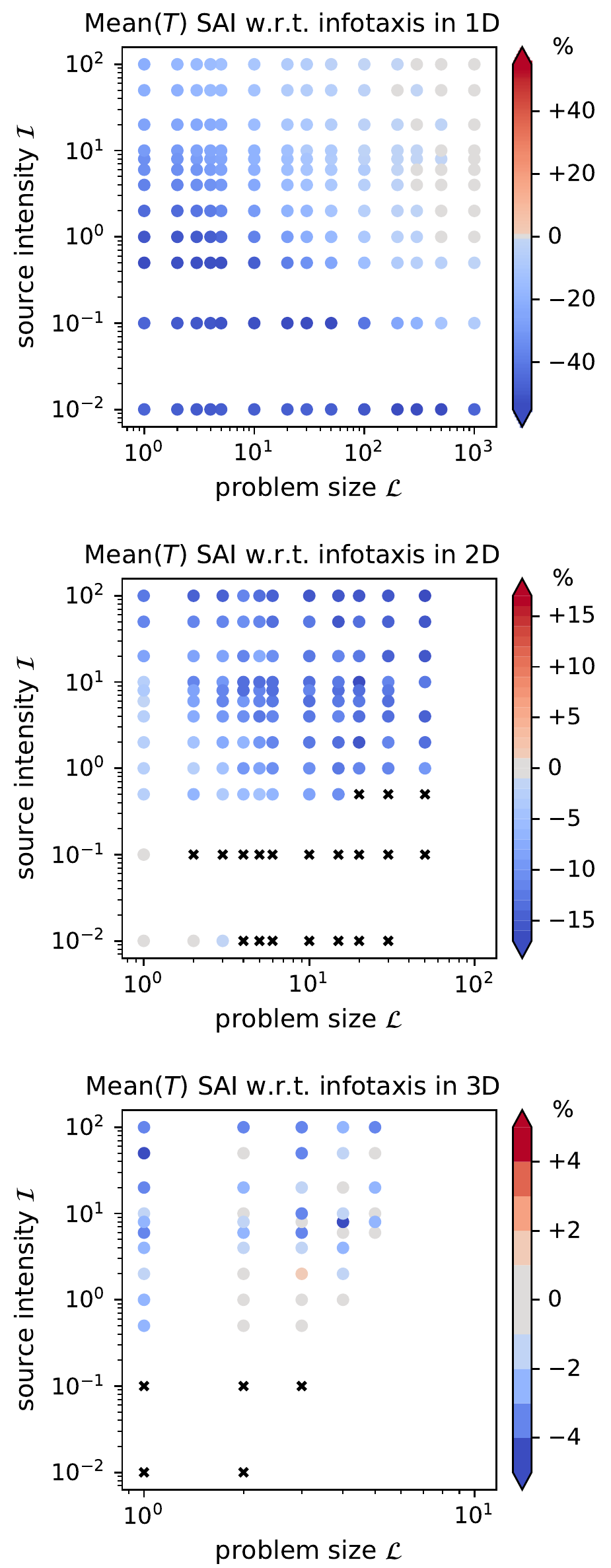}
	\caption{
	Performance of space-aware infotaxis for source tracking in 1D, 2D, and 3D (rows), for a wide range of problem sizes $\mathcal{L}$ and source intensities $\mathcal{I}$: (a) probability of never finding the source and (b) relative difference in the mean number of steps to find the source compared to infotaxis (-10 \% means that SAI finds the source in 10 \% less time than infotaxis). In (b), the black crosses depict cases where $\Pr(\text{failure}) > 10^{-3}$.
	\label{fig:SAI_performance}
	}
\end{figure*}

\subsection{Performance of space-aware infotaxis}

The performance of SAI is plotted in \cref{fig:SAI_performance} as a function of the dimensionless problem size $\mathcal{L}$ and source intensity $\mathcal{I}$ (as defined by \cref{eq:def_dimensionless_parameters}). Video examples of its behavior are provided in Supplementary Material. We show the probability of failure and the relative difference in the mean number of steps between SAI and infotaxis, defined such that a negative value signifies an improvement: for example a relative difference of -10 \% means that SAI finds the source in 10 \% less time than infotaxis, on average. Note that we also tested a version of SAI based on the Euclidean norm (rather than the Manhattan norm used here), and obtained marginally worse performance (see Supplementary Material).

In 1D, SAI never fails and beats infotaxis everywhere by roughly 20 to 50 \% in the lower left quadrant of the ($\mathcal{L}$, $\mathcal{I}$) parameter space, while improvements are more modest in the upper right quadrant. This impressive gain is due to the markedly different behavior of SAI in the absence of hits: while infotaxis tends to go to one end of the domain before turning back and going to the other end, SAI goes back and forth, exploring the domain further each time, as depicted in \cref{appsec:zerohit}. This behavior illustrates the more exploitative nature of SAI compared to infotaxis. 

In 2D, SAI works only in the presence of cues ($\mathcal{I} \gtrsim 1$) but is able to reduce the time to find the source by roughly 5 to 15 \%. In 3D, again SAI works only provided that $\mathcal{I} \gtrsim 1$, and overall performs slightly better that infotaxis  (roughly 2-3 \% improvement overall). We did not evaluate SAI in 4D due to the error bar on infotaxis data that prevents meaningful comparisons. 

The fact that SAI is not reliable for small source intensities is not an important issue, since in this case the search is essentially performed without cues, therefore one could simply use the exhaustive spiral search (which is what infotaxis reduces to in this limit). The lack of reliability of SAI in this case is due to the fact the SAI agent starts with a spiralling trajectory, but after a long period without hits, comes back to the center of the domain and remains trapped there, as illustrated in \cref{appsec:zerohit}.

Importantly, if the source location is known, SAI will direct the agent towards it (in an optimal manner). In contrast, an infotactic agent will be lost in this situation (the entropy being zero, it can not be reduced any further). This explains why SAI is reliable for the high source intensities where infotaxis was shown to fail (cf. \cref{fig:infotaxis_performance}a). 

The distributions of arrival times generated by SAI and by infotaxis exhibit the same features. In particular, the standard deviation is in the order of the mean: for SAI we have $\Std(T) / \Mean(T) \in [0.6, 1.3]$ in 1D, $[0.7, 1.6]$ in 2D, and $[1.0, 2.3]$ in 3D; these values are comparable to those obtained with infotaxis. The standardized distributions of arrival times are similar to those obtained with infotaxis (Supplementary Material).

We now discuss the performance of space-aware infotaxis in the context of prior work. A variant of infotaxis, called entrotraxis, was shown to do better than infotaxis by maximizing the uncertainty on the next observation when the source strength is unknown \cite{Hutchinson2018}. We tested it, and found this is not the case in our setup. Another study suggested to replace the entropy gain by the Bhattacharyya distance \cite{Ristic2016}, but improvement, when existent, is marginal. Finally the $J$ functional of space-aware infotaxis ressembles the free energy functional proposed by Masson \cite{Masson2013}, but the free energy involves a parameter, the temperature, that has to be tuned while $J$ is parameter-free. Besides, we tested our heuristic over the entire parameter space, from 1D to 3D, whereas previous heuristics have only been compared to infotaxis in a limited number of cases. 

\section{Toward the optimal policy: deep reinforcement learning}
\label{sec:RL}

In this section, we show how deep reinforcement learning can be applied to the source-tracking problem. Our goal here is not to perform a complete study of the capabilities of reinforcement learning to solve the source-tracking problem, which could easily be an entire study on its own. Instead, we wish to report on our attempt to solve the source-tracking POMDP using deep reinforcement learning in order to approach at best the optimal policy, as well as to provide a comparison with heuristic policies such as infotaxis and space-aware infotaxis.

\subsection{Reinforcement learning algorithm}

Our reinforcement learning algorithm is based on DQN \cite{Mnih2015}, an extension of traditional Q-learning to deep neural networks. It is adapted here to take advantage of the known model that determines the transitions between a belief state and all its possible successors.

We first recall that (truly) solving the source-tracking problem means finding the optimal policy, as explained in \cref{sec:optimal_policy}. For that, one needs to compute the optimal value function $v^*(s)$, which is the function that satisfies \cref{eq:optimal_Bellman} for all belief states $s$. If belief states could be tabulated, $v^*$ could be obtained by classical dynamic programming using a value iteration algorithm \cite{Bellman1957book,Sutton2018book}. Here, we must resort to an approximate method instead.

We approximate $v^*(s)$ by a parameterized function $\hat{v}(s;\v{w})$ with weight vector $\v{w}$. In practice here, $\hat{v}$ is the function computed by a multi-layer neural network, and $\v{w}$ is the vector containing all ``synaptic'' weights. The Bellman optimality equation for the approximate value function is
\begin{equation}
\label{eq:approximate_optimal_Bellman}
    \hat{v}(s;\v{w}^*) = \min_a \sum_{s'} \Pr(s'|s,a) [ 1 + \hat{v}(s';\v{w}^*)]  \qquad  \forall s \neq s^\Omega
\end{equation}
with $\hat{v}(s^\Omega;\v{w}^*) = 0$, and the problem becomes that of finding the weights $\v{w}^*$ that allow \cref{eq:approximate_optimal_Bellman} to be satisfied ``at best'', that is, the weights that minimize the residual error. This residual error, called the Bellman optimality error, reads
\begin{equation}
\label{eq:loss_definition}
 L(\v{w}) = \mathbb{E}_{s} \left[ \min_a \sum_{s'} \Pr(s'|s,a) [1 + \hat{v}(s';\v{w})] - \hat{v}(s;\v{w}) \right]^2
\end{equation}
where the expectation is taken over belief states $s$ visited when following the policy $\hat{\pi}$ derived from $\hat{v}$ and defined by 
\begin{equation}
\label{eq:approximately_optimal_policy_definition}
    \hat{\pi}(s;\v{w}) = \argmin_a \sum_{s'} \Pr(s'|s,a) [1 + \hat{v}(s';\v{w})].
\end{equation}
The functional $L(\v{w})$ is, in the language of deep neural networks, known as the ``loss function'', and ``training'' the network then refers to the iterative update (through stochastic gradient descent) of the weights $\v{w}$ so as to minimize this loss function. 

It is known that basic training algorithms (such as Q-learning) are unstable or even diverge when nonlinear function approximators are used to represent value functions. This major issue was overcome with DQN, a reinforcement learning algorithm that assembled various stabilization techniques (experience replay, delayed target network) to facilitate convergence, and which capabilities were demonstrated by achieving super-human performance on classic Atari video games such as Pong, Breakout and Space Invaders \cite{Mnih2015}.

Most reinforcement learning algorithms, including Q-learning and DQN, are model-free, and hence rely on the action-value function (also known as the ``Q function''). The source-tracking problem is, however, model-based, because the probability of transitioning from a belief state $s$ to a successor belief state $s'$ is known exactly. This allows us to work directly with the value function $\hat{v}$, and to perform a full backup (that is, to compute the sum over $s'$ in \cref{eq:approximate_optimal_Bellman,eq:loss_definition,eq:approximately_optimal_policy_definition}) rather than a sample backup (based on a single successor belief state randomly sampled).

Our reinforcement learning algorithm is therefore identical to DQN \cite{Mnih2015}, except that the network is trained to approximate the value function rather than the action-value function, and uses full backups rather than sample ones. Further technical information (neural network architecture, hyperparameters, etc.) is provided in Supplementary Material.

\subsection{Performance of (near) optimal policies}

\begin{figure*}
    \flushleft
    \hspace{1.8cm} (a) \\
    \centering
    \includegraphics[width=0.73\linewidth]{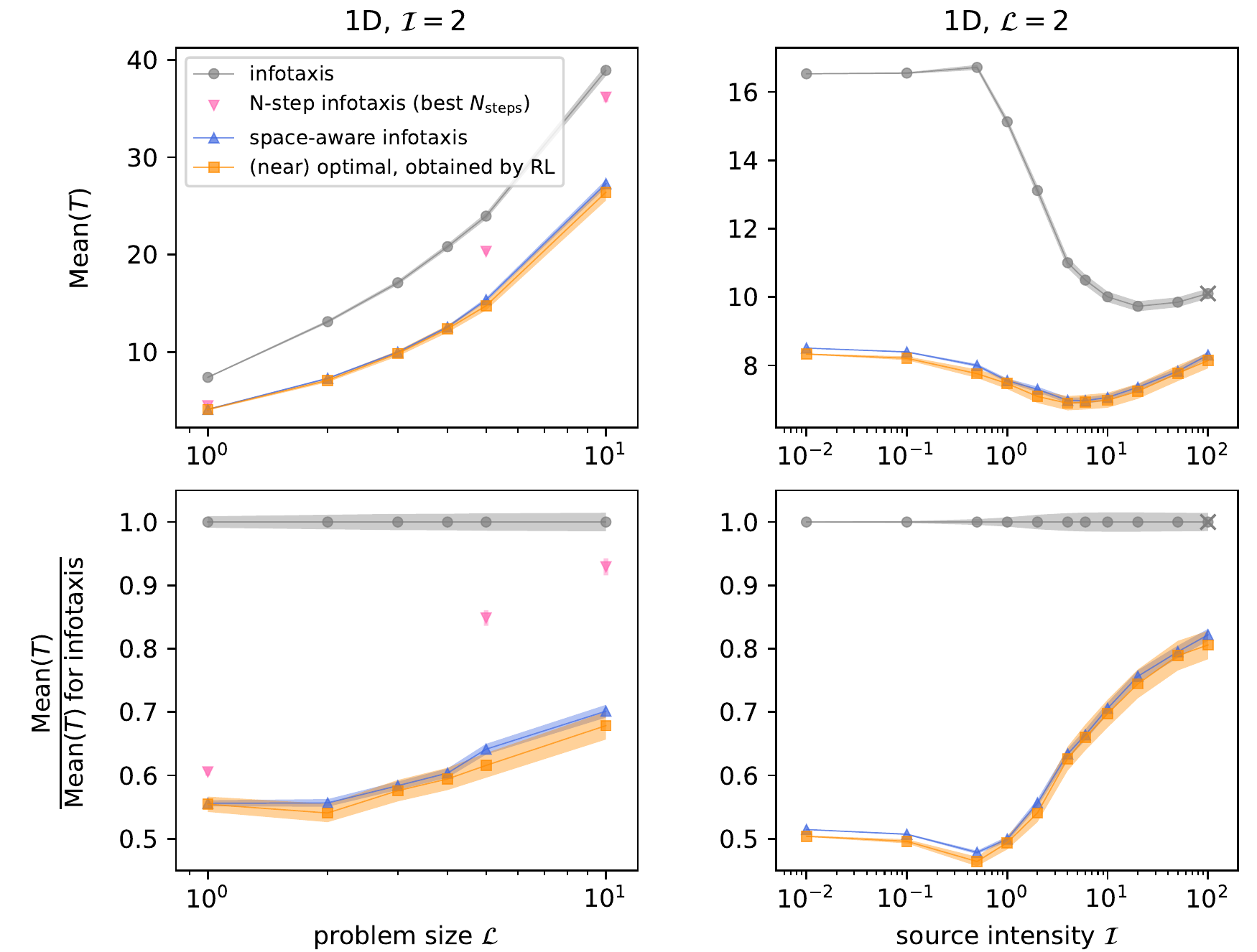} \\[1em]
    \flushleft
    \hspace{1.8cm} (b) \\
    \centering
    \includegraphics[width=0.73\linewidth]{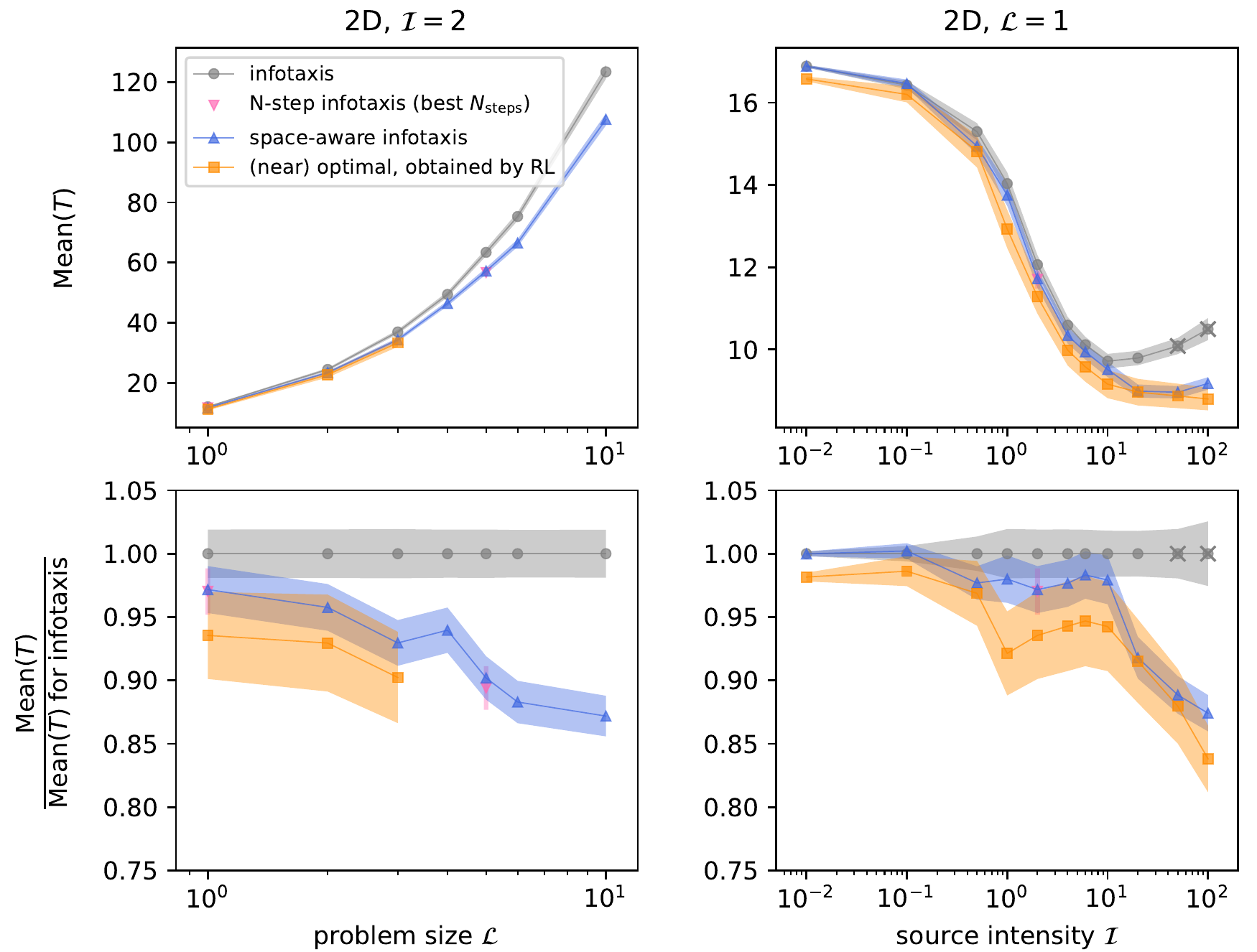}
	\caption{
	Performance of (near) optimal policies obtained by deep reinforcement learning (RL) in (a) 1D and (b) 2D: mean number of steps to find the source as a function of the problem size $\mathcal{L}$ (left) and the source intensity $\mathcal{I}$ (right). All RL policies have $\Pr(\text{failure}) < 10^{-6}$. For comparison we also show the performance of infotaxis, space-aware infotaxis and N-step infotaxis (when available). Crosses indicate that $\Pr(\text{failure}) > 10^{-3}$. Shaded areas show 95 \% confidence intervals.
	\label{fig:RL_results}
	}
\end{figure*}

In order to obtain the best possible policy, we kept on increasing the size (width and depth) of the neural network until the performance of the learned policy stopped improving. This is a good indication (though not a proof) that the learned policy is optimal or very close to optimal. Due to the high cost of training and of ensuring near-optimality, we trained policies only for 1D and 2D domains. Importantly, neural networks were trained from scratch (random initial weights) and all the results we present are very reproducible (e.g., for different random initializations). Several videos that illustrate the behavior of these (near) optimal policies are available in Supplementary Material. 

The trajectories obtained in the absence of hits ($\mathcal{I} \rightarrow 0$) are shown in \cref{appsec:zerohit}. In 1D, the (near) optimal trajectory is qualitatively similar (but not identical) to the one generated by space-aware infotaxis with a 'back and forth' motion extending further each time.  In 2D, the (near) optimal trajectory is essentially an exhaustive spiral, as we obtained with infotaxis and space-aware infotaxis. Its geometry is however different, since it approximates a continuous Archimedean spiral defined with a Euclidean norm, rather than one defined with Chebyshev norm (as was the case with infotaxis and space-aware infotaxis). Repeated trainings produced slightly different trajectories, though all looked very similar to the one shown in \cref{appsec:zerohit}. In continuous space, the Euclidean Archimedean spiral can be proven to be the optimal path for $\mathcal{I} \rightarrow 0$ using the fact that $p_0$ (the initial source probability distribution) is a monotonically decreasing function of the initial distance to the agent. Infotaxis has been shown to follow such a spiral in continuous space \cite{Barbieri2011}, hence the differences we observe here between (space-aware) infotaxis and the reinforcement learning policy are only due to our discrete setting.

The (near) optimal performance of learned policies is reported in \cref{fig:RL_results} for searches in 1D and 2D with varying source intensities $\mathcal{I}$, and, to some extent, varying problem sizes $\mathcal{L}$ (these dimensionless parameters are defined by \cref{eq:def_dimensionless_parameters}). Importantly all these learned policies have $\Pr(\text{failure}) < 10^{-6}$. For comparison we also show the performance of infotaxis, N-step infotaxis and space-aware infotaxis on the same figure. In 1D, the mean time to find the source can be reduced by roughly 20 to 50 \% compared to infotaxis. In 2D, this reduction is roughly around 5-10 \%, and up to 16 \%. 

The most surprising result is maybe the remarkable performance of space-aware infotaxis: for all cases considered here, the performance attained with our heuristic is very close to the (near) optimal one obtained by deep reinforcement learning. It will be very interesting to evaluate to which extent this result generalizes to larger domain sizes and to 3D.

\section{Discussion: why does infotaxis work so well?}
\label{sec:discussion}

We have shown in this paper that while infotaxis is generally suboptimal, it performs remarkably well in a wide range of searching conditions. This is somewhat suprising, because the entropy of a belief state is not linearly related to its optimal value (\cref{sec:infotaxis}). In this section we discuss possible explanations to this good performance.

A theoretical argument was proposed in ref.~\cite{Vergassola2007}, where the authors show that given a probability distribution $p(\v{x})$ with entropy $H$, the expected optimal search time satisfies $\mathbb{E}_{\pi^*} [T | p_0 = p(\v{x})] \geqslant 2^{H - 1}$, and conclude that reducing entropy is a necessary condition to an efficient search. This bound is, according to the authors, a lower bound because it assumes the agent can jump to any cell for a unit cost. This result, however, is derived assuming a frozen distribution $p(\v{x})$ and therefore only holds in the absence of cues gathered along the search ($\mathcal{I} \rightarrow 0$). This is the same assumption we actually used to derive our upper bound in \cref{eq:upper_bound} based on an exhaustive search, and also for our estimate of the remaining time for space-aware infotaxis in \cref{eq:space_aware_infotaxis_effect_of_entropy}. It can be easily verified, from our data but also from Fig.~2c in ref.~\cite{Vergassola2007}, that it is not a valid lower bound for a search with cues. Besides, based on this argument, one should seek to minimize the expectation of $2^{H-1}$ at the next step, rather than that of $H$ (since these expressions are not linearly related, their expectations may rank actions differently). We actually tested this idea, and found that it does not perform as well as infotaxis.

We believe that the correct theoretical argument lies in the concavity of the optimal value function. This argument was provided in early artificial intelligence papers on POMDPs \cite{Cassandra1996,Kaelbling1998}, where the idea of entropy minimization was proposed for robotic navigation. First recall that solving the source-tracking problem is equivalent to computing its optimal value function $v^*$. The optimal policy is then to choose the action that minimizes the expected optimal value at the next step. It can be mathematically proven that the optimal value function is a concave function of the belief state, in our case $p(\v{x})$ \cite{Sondik1971thesis,Smallwood1973}. The concavity of the optimal value function implies that the lowest values correspond to belief states located in the corners of the belief space, where $p(\v{x}) = \delta(\v{x} - \v{x}')$: these are also the belief states with the lowest entropy.

Quoting Kaelbling et al. \cite{Kaelbling1998} for an intuitive explanation: ``The [concavity] of the optimal value function makes intuitive sense when we think about the value of belief states. States that are in the 'middle' of the belief space have high entropy -- the agent is very uncertain about the real underlying state of the world. In such belief states, the agent cannot select actions very appropriately [...]. In low-entropy belief states, which are near the corners of the simplex, the agent can take actions more likely to be appropriate for the current state of the world [...]. This has some connection to the notion of 'value of information', where an agent can incur a cost to move it from a high-entropy to a low-entropy state; this is only worthwhile when the value of the information (the difference in value between the two states) exceeds the cost of gaining the information.'' 

As we have seen in this paper, beating infotaxis seems to be increasingly difficult as the dimensionality increases. While this could be inherent to the greater difficulty of finding good approximate solutions in higher dimensions, this result is well explained by the above argument: higher dimensionality implies larger uncertainty ($H \sim n \log N$, with $n$ the dimension and $N$ the linear domain size), meaning that optimal actions are those which help disambiguate the true source location (those which reduce entropy) for a larger fraction of the entire search. 

\section{Conclusion}
\label{sec:conclusion}

In this paper we demonstrated that infotaxis is generally suboptimal and can be beaten by other heuristics or by reinforcement learning, in particular for searches in lower dimensional spaces (and so even if an anticipation over multiple steps rather than one step is allowed). Yet, it remains a strong contender because of its generality: indeed infotaxis performs well over a huge range of parameters without any tuning. More precisely, we have shown that (i) the probability of failing to find the source is negligible (infotaxis is \emph{reliable}), (ii) the mean search time scales with physical parameters as one would expect for the optimal policy (infotaxis is \emph{efficient}), and (iii) the tail of the distribution of search times decays faster than any power law, though subexponentially (infotaxis is not plagued by large fluctuations and hence is \emph{safe}). 
Finally, we have shown that infotaxis can be made more efficient if the uncertainty measure (entropy) is balanced by a distance measure. We called this parameter-free policy ``space-aware infotaxis''. Overall, this new heuristic reduces the mean time to find the source by 10-50 \% in 1D, 5-15 \% in 2D, and 2-3 \% in 3D.

To provide a firmer answer to the question of how good infotaxis is, one needs to compare it with the optimal policy. The nature and the size of the source-tracking problem do not allow the computation of exact solutions, however finding approximately optimal solutions is a task well-suited for deep reinforcement learning. Our learning algorithm is a model-based version of DQN \cite{Mnih2015} where the value function is approximated by a deep neural network. We used large neural networks in order to approach at best the optimal policy. Because of the computational cost, we only trained policies in 1D and 2D. Our results demonstrate that policies superior to infotaxis can be learned from scratch purely from experience. We found that, for the cases considered, the mean time to find the source can be reduced by 20-50 \% in 1D and 5-15 \% in 2D compared to infotaxis. Almost identical improvements were obtained by our space-aware infotaxis in those cases. Overall, this strongly suggests that (i) while infotaxis is vastly suboptimal in 1D, the margin of improvement toward the optimal policy gets tighter as the dimensionality increases and (ii) space-aware infotaxis is probably a very good approximation of the optimal policy.

The source-tracking problem was inspired by olfactory searches performed by moths, which use pheromones to locate their mates \cite{Hansson1999book,Murlis1992}. A similar navigation task is also faced by crustaceans looking for food on the sea floor \cite{Koehl2001,Reidenbach2011}. It is not plausible that these animals perform infotaxis as presented here \cite{Carde2021}: infotaxis involves complex computations (due to the 1-step lookahead into possible futures) and relies on unrealistic assumptions (physical space representation, perfect memory, model knowledge). Developing heuristic approximations of infotaxis with much weaker requirements is possible \cite{Masson2013}, and it will be interesting to see whether such memory-based strategies can be discovered by minimal recurrent neural networks through reinforcement learning, as recently achieved for chemotaxis \cite{Hartl2021}. 

\begin{figure*}
 \includegraphics[width=0.99\linewidth]{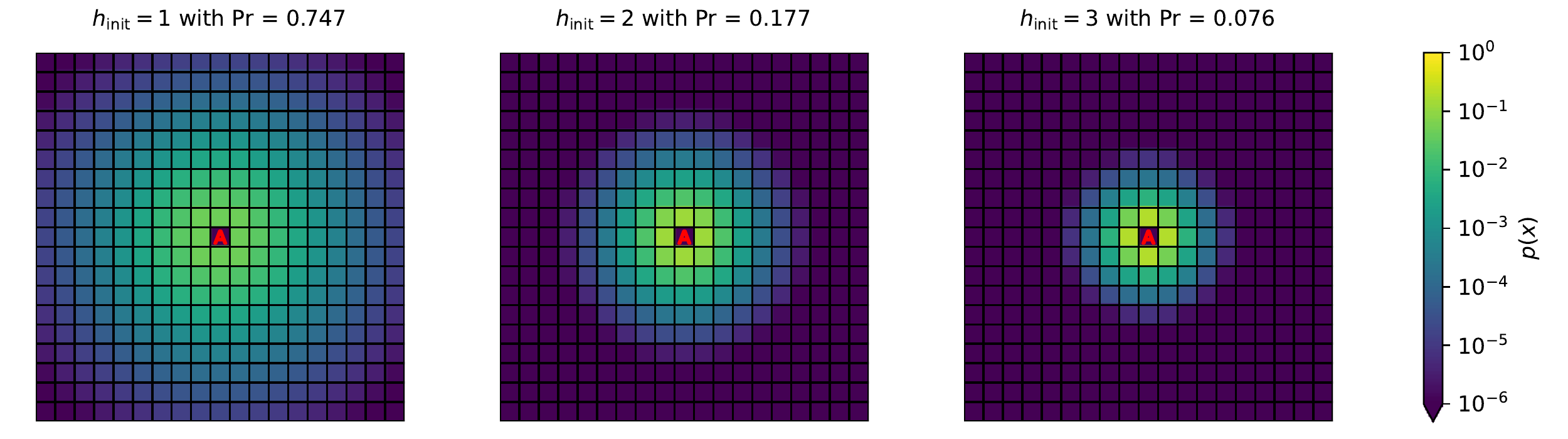}
 \caption{Initialization procedure: a set of initial beliefs $p_0$, here in 2D for $\mathcal{L}=1$ and $\mathcal{I} = 2$, with the corresponding values of the initial hit and probabilities of occurring shown on top. ``A'' indicates the agent's position, at the center of the domain.}
 \label{fig:illustration_initial_set_of_beliefs}
\end{figure*}

Reinforcement learning has recently emerged as a powerful tool for finding nontrival solutions to complex navigation problems in turbulent flows \cite{Reddy2016a,Reddy2018,Colabrese2017,Gustavsson2017,Alageshan2020}, and provides a complementary approach to physics-based heuristics \cite{Monthiller2021arxiv}. 
While all these navigation problems are POMDPs, they have never been explicitely framed as such and hence never exploited existing mathematical results and algorithms developed for this class of problems. In addition, learned policies are most often memoryless, while it is clear that navigation strategies shaped by Darwinian evolution rely on memory, either in a physical (molecular) form for single cells (e.g., bacteria performing chemotaxis \cite{Macnab1972,Lan2016}) or in an abstract one for higher organisms with cognitive capabilities (from the worm \textit{C. elegans} to humans).

We believe that combining ideas from infotaxis (as a viable alternative to physical gradient climbing), Bayesian inference (for memory encoding) and POMDPs (which provide the mathematical framework) while leveraging the power of deep reinforcement learning is a promising route for uncovering search and navigation strategies used by living organisms and for adapting those to robotic applications. Through this paper, we made a first step in this direction by bringing together the relevant concepts and methods, and we hope to trigger further interest from all relevant communities on this highly interdisciplinary topic.

\vspace{1em}
This study has been performed with the open-source code OTTO, which is freely available on GitHub at \url{http://github.com/C0PEP0D/otto} \cite{OTTO}. The data that support the findings of this study are openly available on Zenodo at \url{https://doi.org/10.5281/zenodo.6125391} \cite{dataset}. 

\begin{acknowledgments}
This project has received funding from the European Research Council (ERC) under the European Union's Horizon 2020 research and innovation programme (grant agreement No 834238). 
Centre de Calcul Intensif d'Aix-Marseille is acknowledged for granting access to its high performance computing resources.
\end{acknowledgments}

\appendix 

\section{Initialization protocol}
\label{appsec:initialization_protocol}

\begin{figure*}
    \centering
    \includegraphics[width=0.99\linewidth]{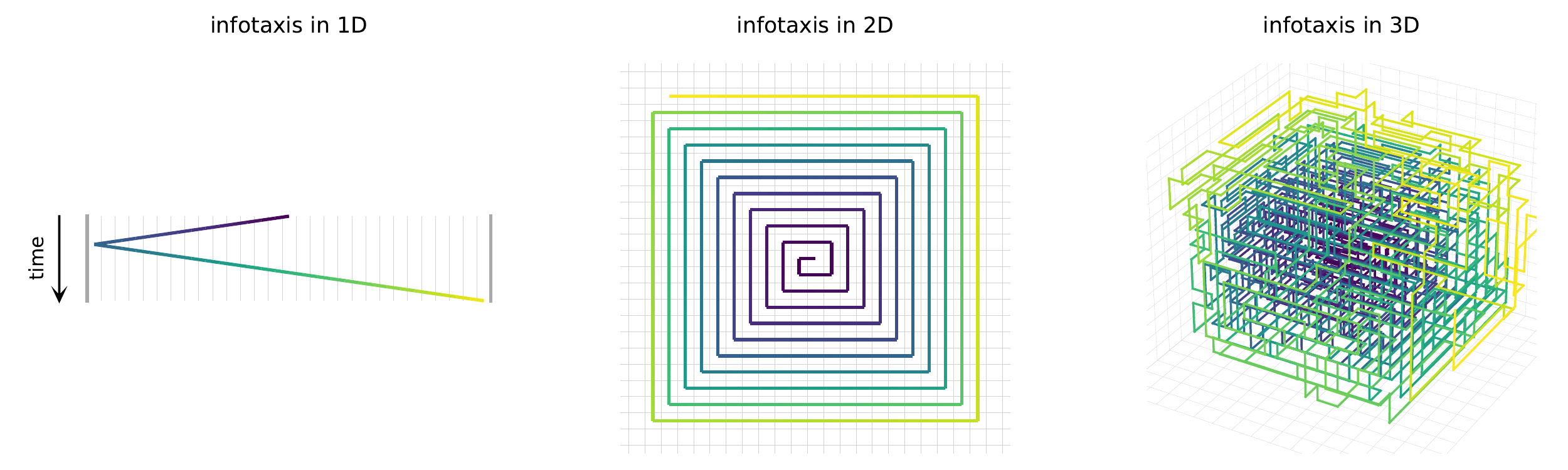} \\[1em]
    \includegraphics[width=0.99\linewidth]{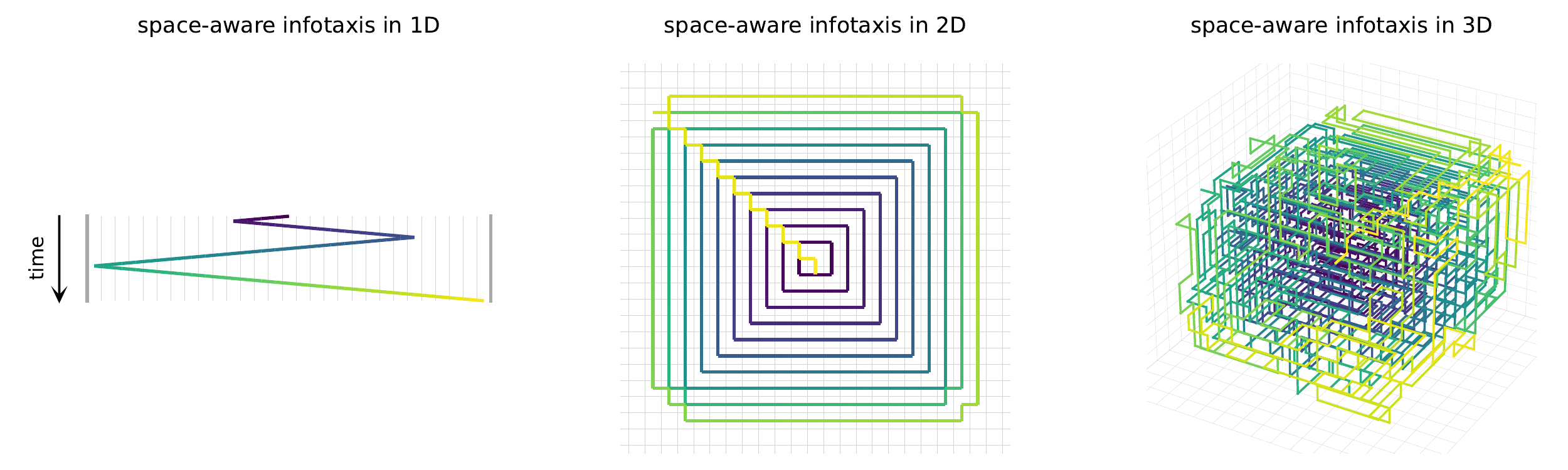} \\[1em]
    \includegraphics[width=0.99\linewidth]{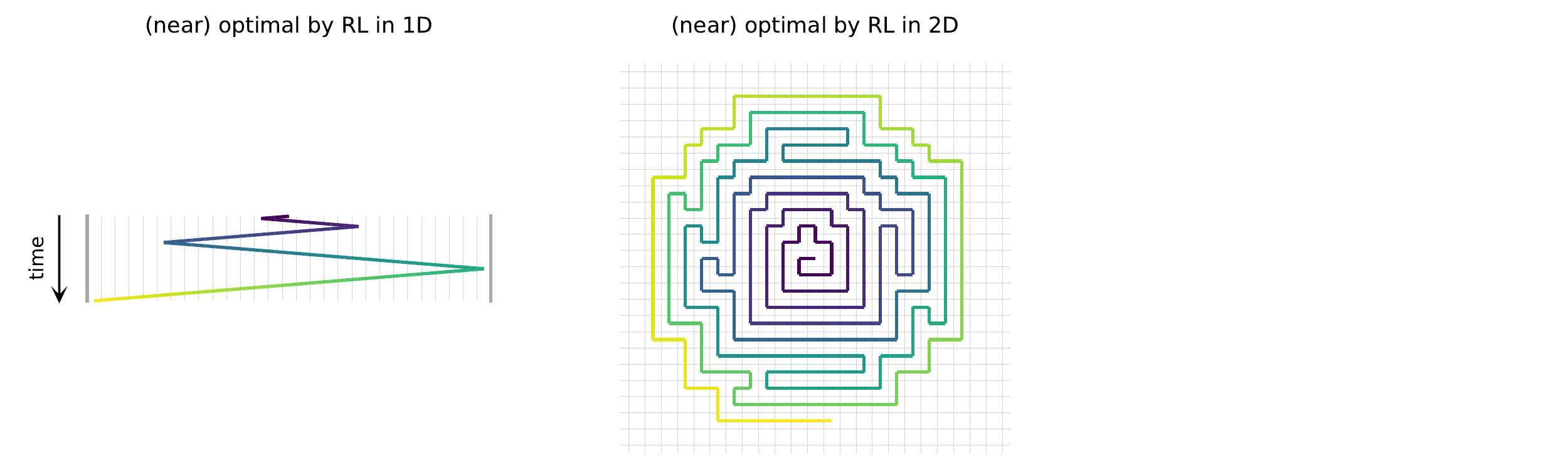} \\
	\caption{
    Trajectories in the absence of cues (succession of zero hits in the limit $\mathcal{I} \rightarrow 0$) for various policies: infotaxis (top), space-aware infotaxis (middle), near-optimal by deep reinforcement learning (bottom). Time is color-coded from blue (start) to yellow (end). The grid is shown in grey. 
	\label{fig:zerohit_trajectories}
	}
\end{figure*}

The physical model used in the source-tracking problem assumes an infinite, open domain. Because tracking can only start once the searcher knows that a source exists (that is, after a detection event has occurred), we propose to start the search with a nonzero hit and give a criteria to define a numerical domain ``large enough'' to mimic an open domain.

To simplify the derivation we will assume in the following a continuous space. Consider a spherical coordinates system (in $n$ dimensions) centered on the agent. We denote $\Pr(r) \, \d r$ the probability of the source being in a spherical shell of radius $r$ and thickness $\d r$. Before the search starts, we assume a uniform prior over the entire space and excluding the agent volume:
\begin{equation}
 \Pr(r) = 
\begin{dcases}
    0 & \text{for $r \in [0,a]$} \\ 
    c S_n(r) & \text{for $r>a$}
\end{dcases}
\end{equation}
where $a$ is the agent radius, $c$ is a constant and $S_n(r)$ is the surface area of the n-ball of radius $r$
\begin{equation}
 S_n(r) = \frac{2 \upi^{n/2} r^{n-1}}{\Gamma(n/2)}
\end{equation}
with $\Gamma$ the gamma function.
After a initial hit $h_{\text{init}}$, using Bayes' rule, we can write
\begin{equation}
    \Pr(r | h_{\text{init}}) = \frac{\Pr(h_{\text{init}} | r) \Pr(r)}{\Pr(h_{\text{init}})} \quad \text{for $r>a$}
\end{equation}
with the normalization constant given by
\begin{equation}
 \Pr(h_{\text{init}}) = \int_{a}^\infty \Pr(h_{\text{init}} | r) \Pr(r) \id r.
\end{equation}

For $h_{\text{init}} \neq 0$, $\Pr(r | h_{\text{init}})$ decays exponentially fast with $r$ for large $r$. We can define a cut-off radius $R_c$ such that the probability of the source being outside a ball of radius $R_c$ is at most $\varepsilon_{\text{out}}$ for any $h_{\text{init}}$ (with $\varepsilon_{\text{out}}$ small). This reads
\begin{equation}
 R_c = \max_{h_{\text{init}} \neq 0} r_c(h_{\text{init}})
\end{equation}
where $r_c(h_{\text{init}})$ is implicitely defined by
\begin{equation}
 \int_{a}^{r_c(h_{\text{init}})} \Pr(r | h_{\text{init}}) \id r = 1 - \varepsilon_{\text{out}}.
\end{equation}

The initialization procedure for the search is therefore as follows.
\begin{enumerate}
    \item The initial hit $h_{\text{init}}$ is drawn at the beginning of each episode from the corresponding probability distribution (excluding zero hit):
    \begin{equation}
    \label{eq:non_zero_hit_proba}
    \Pr(h_{\text{init}} | h_{\text{init}} \neq 0) = 
    \begin{dcases}
    0 & \text{if $h_{\text{init}} = 0$} \\ 
    \Pr(h_{\text{init}}) / Z & \text{otherwise}
    \end{dcases}
    \end{equation}
    with $Z$ the normalization constant such that $\sum_{h_{\text{init}}} \Pr(h_{\text{init}} | h_{\text{init}} \neq 0) = 1$.
    \item The linear size of the grid, $N$, is set to 
    \begin{equation}
     N=2 \ceil(R_c) + 1.
    \end{equation}
    \item The source distribution is set to uniform: $p_{\text{init}}(\v{x}) = 1/N^n$.
    \item The agent's position $\v{x}^a_0$ is set to the center of the domain.
    \item The initial source distribution $p_0$ is computed using $p_0 = \Bayes(p_{\text{init}}(\v{x}), \v{x}^a_{0}, (\bar{F}, h_{\text{init}}))$, and the initial belief state is then $s_0=[\v{x}^a_0, p_0(\v{x})]$.
\end{enumerate}
Following this procedure, the set of initial beliefs $p_0$ is the set generated by $h_{\text{init}}=\{1, 2, \dots\}$ and their probabilities of occurring are given by \cref{eq:non_zero_hit_proba}. An example of such a set is shown in \cref{fig:illustration_initial_set_of_beliefs}. We used $\varepsilon_{\text{out}} = 10^{-3}$ in all our simulations. As a rule of thumb, this gives $N \approx 15 \mathcal{L}$. More information on how $N$ depends on the parameters is provided in Supplementary Material.

\section{Policy evaluation}
\label{appsec:policy_evaluation}

Policy evaluation is performed by generating a large number of episodes and computing the resulting distribution of arrival times $T$, denoted here $f(T)$.

The convergence of $f(T)$ with the number of episodes can be vastly improved by realizing that $p(\v{x})$, interpreted as the agent's belief in the context of decision-making, is also the true (in the Bayesian sense) probability distribution of sources that could have generated the sequence of observations. In this probabilistic approach, each episode can be continued until the probability of having found the source is equal to one (within numerical accuracy $\varepsilon_{\text{stop}} = 10^{-6}$) or until the agent is stuck in an infinite loop, and the hits are drawn at each step according to the distribution:
\begin{equation}
    \label{eq:draw_hit_distributed_source}
    \Pr(h | \v{x}^a) = \sum_{\v{x}} \Pr(h | \v{x}^a, \v{x}) p(\v{x})
\end{equation}
such that episodes can be generated independently of the true source location $\v{x}^s$. We refer to this alternative framework as ``hybrid Bayesian/Monte-Carlo'' (as opposed to ``full Monte-Carlo'').
A video illustrating how the search proceeds in this framework, together with numerical proofs of its correctness and efficiency, are provided in Supplemental Material. This approach is particularly advantageous to sample rare events (such as failing to find the source) and more generally to sample heavy-tailed distributions (as is $f(T)$). 

The number of episodes is chosen such that the mean of the distribution is well-converged (95\% confidence interval less than $\pm 2 \%$). Typically we use at least 16000 episodes in 1D, 6400 episodes in 2D, and 25600 episodes in 3D. In 4D, we determined the mean with less accuracy due to the high cost of the simulations: we used typically 4096 episodes (yielding a 95\% confidence interval less than $\pm 10 \%$) and down to 1024 episodes for the most expensive cases.

\section{Zero-hit trajectories}
\label{appsec:zerohit}

In the limit of a vanishing source intensity ($\mathcal{I} \rightarrow 0$), the search is performed without cues and the trajectories are deterministic, because they are generated by a succession of zero hits. These trajectories are shown in \cref{fig:zerohit_trajectories} for infotaxis, space-aware infotaxis, and the near-optimal policy obtained by deep reinforcement learning. They have been calculated using $\mathcal{I}=10^{-6}$ and $\mathcal{L}=2$ in 1D, $\mathcal{L}=1.5$ in 2D, and $\mathcal{L}=1$ in 3D.

\bibliography{biblio.bib}

\end{document}